\documentclass[letterpaper, 11pt]{article}
\pdfoutput=1

\usepackage[nocompress]{cite}
\usepackage{jheppub}

\usepackage{graphicx}
\usepackage{epstopdf}
\usepackage{amsmath, amssymb}

\usepackage{bm}
\usepackage{bbold}
\usepackage{caption}
\usepackage{subcaption}

\newcommand{\be}{\begin{eqnarray}}
\newcommand{\ee}{\end{eqnarray}}
\newcommand{\nn}{\nonumber}
\newcommand{\bn}{\begin{enumerate}}
\newcommand{\en}{\end{enumerate}}

\def\IC{\mathbb{C}}

\def\IP{\mathbb{P}}

\def\IZ{\mathbb{Z}}

\def\CB{{\cal B}}

\def\CE{{\cal E}}
\def\CF{{\cal F}}

\def\CI{{\cal I}}

\def\CL{{\cal L}}
\def\CM{{\cal M}}
\def\CN{{\cal N}}
\def\CO{{\cal O}}

\def\CS{{\cal S}}
\def\CT{{\cal T}}

\def\a{\alpha}
\def\b{\beta}

\def\e{\epsilon}

\def\k{\kappa}

\def\s{\sigma}

\def\G{\Gamma}

\def\half{\frac{1}{2}}

\def\vev#1{\langle #1 \rangle}

\def\Tr{{\rm Tr}}
\def\tr{{\rm Tr}}

\def\ft{\mathfrak{t}}
\def\PE{\textrm{PE}}

\def\vec#1{\bm{#1}}

\newcommand{\bea}{\begin{eqnarray}}
\newcommand{\eea}{\end{eqnarray}}

\def\IC{\mathbb{C}}

\def\IP{\mathbb{P}}

\def\IZ{\mathbb{Z}}

\def\CB{{\cal B}}

\def\CE{{\cal E}}
\def\CF{{\cal F}}

\def\CI{{\cal I}}

\def\CL{{\cal L}}
\def\CM{{\cal M}}
\def\CN{{\cal N}}
\def\CO{{\cal O}}

\def\CS{{\cal S}}
\def\CT{{\cal T}}

\def\a{\alpha}
\def\b{\beta}

\def\e{\epsilon}

\def\k{\kappa}

\def\s{\sigma}

\def\G{\Gamma}

\def\half{\frac{1}{2}}

\def\vev#1{\langle #1 \rangle}

\def\Tr{{\rm Tr}}
\def\tr{{\rm Tr}}

\def\PE{{\rm PE}}

\title{$\CN=1$ Deformations and RG Flows of $\CN=2$ SCFTs}

\author[a,b]{Kazunobu Maruyoshi}
\author[c]{and Jaewon Song}
\affiliation[a]{Department of Physics, Imperial College London\\ Blackett Laboratory, Prince Concert Road, South Kensington, London, SW7 2AZ, UK}
\affiliation[b]{Faculty of Science and Technology, Seikei University\\ 3-3-1 Kichijoji-Kitamachi, Musashino-shi, Tokyo, 180-8633, Japan}
\affiliation[c]{Department of Physics, University of California, San Diego \\La Jolla, CA 92093, USA}
\emailAdd{maruyoshi@st.seikei.ac.jp}
\emailAdd{jsong@physics.ucsd.edu}

\abstract
{ We study certain $\CN=1$ preserving deformations of four-dimensional $\CN=2$ superconformal field theories (SCFTs)
  with non-abelian flavor symmetry. The deformation is described by adding an $\CN=1$ chiral multiplet transforming in the adjoint representation of the flavor symmetry with a superpotential coupling, and giving a nilpotent vacuum expectation value to the chiral multiplet
  which breaks the flavor symmetry.
  This triggers a renormalization group flow to an infrared SCFT.
  Remarkably, we find classes of theories flow to enhanced $\CN=2$ supersymmetric fixed points in the infrared under the deformation.
  They include generalized Argyres-Douglas theories and rank-one SCFTs with non-abelian flavor symmetries.  
Most notably, we find renormalization group flows from the deformed conformal SQCDs to the $(A_1, A_n)$ Argyres-Douglas theories. 
From these ``Lagrangian descriptions," we compute the full superconformal indices of the $(A_1, A_n)$ theories 
and find agreements with the previous results.
Furthermore, we study the cases, including the $T_N$ and $R_{0,N}$ theories of class $\CS$ and some of rank-one SCFTs, where the deformation gives genuine $\CN=1$ fixed points.
}

\preprint{IMPERIAL-TP-16-KM-03}

\begin{document}
\maketitle

\section{Introduction}
  Renormalization group (RG) flow of a supersymmetric theory preserving its full or part of supersymmetries is 
  a quite non-trivial but remarkably tractable phenomenon,
  thanks to various techniques developed in the past several decades \cite{Seiberg:1994pq, Leigh:1995ep, Intriligator:2003jj}. 
  Physics at the infrared fixed point is described by a superconformal field theory (SCFT), thus the RG flow defines the SCFT.
  
  In this paper, we study the deformation of four-dimensional $\CN=2$ SCFTs which preserves at least $\CN=1$ supersymmetry
  and the RG flow caused by this.\footnote{Our deformation does not belong to the one classified in \cite{Cordova:2016xhm}, 
    since we add extra chiral multiplets to the theory.}
  The deformation is described as follows:
  suppose we have an $\CN=2$ SCFT, $\CT$, with a non-Abelian (semi-simple) flavor symmetry $F$, then
  \begin{itemize}
  \item add an $\CN=1$ chiral multiplet $M$ transforming in the adjoint representation of $F$ via the superpotential coupling
           with the moment map operator $\mu$ of $F$ given by
           \be
           	W = \tr M \mu \ ,
           \ee           
  \item then give a nilpotent vacuum expectation value (vev) to the chiral multiplet $M$ 
  	\be
	 \vev{M} = \rho(\s^+) \ ,
	\ee
	where $\s^+ = \s_1 + i \s_2$ and $\sigma_i$ are the Pauli matrices. 
  \end{itemize}
  The nilpotent vev is specified by the embedding $\rho$: $\mathfrak{su}(2) \rightarrow \mathfrak{f}$, where $\mathfrak{f}$ is the Lie algebra of $F$.
  When $F=SU(N)$, the embedding is classified by the partition of $N$.
  This procedure defines the theory $\CT_{{\rm IR}}[\CT, \rho]$ in the IR fixed point.
  
  Among the embeddings (or the partitions), we focus here on the flow by the principal embedding. 
  The principal embedding is the one $\rho$ such that $\mathfrak{f}$ is decomposed into rank-$\mathfrak{f}$ irreducible representations.
  Thus this breaks the flavor symmetry $F$ completely and leaves the theory with only $U(1)_\CF$ global symmetry 
  coming from the Cartan of the $SU(2)_R \times U(1)_r$ symmetry of $\CT$ (and Abelian factors of the original flavor symmetry). 
  This $U(1)_\CF$ symmetry can mix with $R$-symmetry in the IR, 
  and the superconformal R-symmetry is determined via $a$-maximization \cite{Intriligator:2003jj}. 
  Remarkably we observe in various examples that 
  {\it the fixed point theory due to the principal embedding has an enhanced $\CN=2$ supersymmetry.}
  Along the RG flow the scaling dimensions of some of the chiral operators hit the unitarity bound and get decoupled.
  Therefore the $\CN=2$ supersymmetry in the IR cannot be the original UV one, rather it is an accidental symmetry.
  
  However it is not always the case that the principal embedding leads to the enhancement of the supersymmetry in the IR. 
  While it would be interesting to figure out the physical mechanism and even the necessary condition for $\CT$ to have this enhancement,
  we only list the theories with the enhancement in this paper.
  Indeed, we check the IR enhancement when $\CT$ is the following SCFT:
    \begin{itemize}
    \item the so-called $(A_1, D_k)$ \cite{Cecotti:2010fi,Bonelli:2011aa,Argyres:2012fu} 
    and $(I_{N, Nm+1}, F)$ theories \cite{Xie:2012hs} of Argyres-Douglas type
    \item the rank-one SCFTs $H_1$, $H_2$, $D_4$, $E_6$, $E_7$ and $E_8$ \cite{Seiberg:1994aj, Argyres:1995xn,Minahan:1996fg,Minahan:1996cj}.
    \item Conformal SQCDs: $SU(N)$ gauge theory with $2N$ fundamental hypermultiplets and $Sp(N)$ gauge theory with $4N+4$ fundamental half-hypermultiplets \cite{Argyres:1995wt, Argyres:1995fw}.
    \end{itemize}
  For the first case, the IR theories are $(A_1, A_{k-1})$ and $(A_{N-1}, A_{Nm+N})$ theories of Argyres-Douglas type respectively.
  For the second case, one always gets the simplest $\CN=2$ SCFT, $H_0$ \cite{Argyres:1995jj}.
  For the last case, the IR theories are the $(A_1, A_{2N-1})$ and $(A_1, A_{2N})$ theories respectively.
  
  Our discussion does not depend on whether the theory $\CT$ admins a Lagrangian description or not. 
  Most examples we study do not have (known) Lagrangian descriptions.  
  But special cases are when $\CT$ is the $D_4$ theory, and the conformal SQCDs,
  which have Lagrangian realizations.   
  Therefore the deformation due to the principal embedding can be analyzed in a Lagrangian level. 
  This leads in the $D_4$ case to the theory studied recently in \cite{Maruyoshi:2016tqk}, which flows to $H_0$ in the IR.
  The $SU(N)$ SQCD with $2N$ flavors has the $SU(2N) \times U(1)$ flavor symmetry. 
  We break the $SU(2N)$ part of the global symmetry by the deformation. 
  This triggers a flow to the $(A_1, A_{2N-1})$ theory, 
  which has the $U(1)$ global symmetry for $N>2$ and $SU(2)$ for $N=2$.  
  For the $Sp(N)$ SQCD, the deformation breaks the $SO(4N+4)$ flavor symmetry completely. 
  Note that when $SU(2) = Sp(1)$, depending on breaking the $SU(4)$ subgroup of the flavor symmetry or the entire $SO(8)$ flavor symmetry, we obtain flow to the $H_1 = (A_1, A_3)$ or the $H_0 = (A_1, A_2)$ theory.
  
  These Lagrangian theories open up a way to compute the superconformal index \cite{Kinney:2005ej, Romelsberger:2005eg} 
  of the IR theory in full generality,
  namely with $p$, $q$ and $t$ fugacities.
  The only nontrivial issue is the existence of the decoupled chiral multiplets along the RG flow whose contributions are subtracted by hand,
  as in \cite{Kutasov:2003iy} for the central charge computation.
  This was demonstrated in \cite{Maruyoshi:2016tqk} for the $H_0$ theory.
  We give the expressions for the full superconformal indices of the $(A_1, A_k)$ theories here.
  
  On the other hand, we find that the following theories flow to genuine $\CN=1$ SCFTs:
    \begin{itemize}
    \item A number of rank-one SCFTs \cite{Argyres:2007tq, Argyres:2015ffa,Argyres:2015gha,Chacaltana:2016shw, Argyres:2016xua} 
     do not belong to the $H_{1, 2}, D_4, E_{6, 7, 8}$ series,
    \item $T_N$ theory and $R_{0,N}$ theory in class $\CS$ \cite{Gaiotto:2009we, Chacaltana:2010ks},
    \item $\CN=4$ $SU(2)$ super Yang-Mills theory,
    \end{itemize}
  as can be seen by irrational central charges.\footnote{Let us argue that any $\CN=2$ SCFT should have rational central charges. Assuming any $\CN=2$ SCFTs has a suitable Seiberg-Witten (SW) geometry describing its Coulomb branch, the central charges can be obtained from the SW curve \cite{Shapere:2008zf}. The SW geometry is given by an algebraic curve and a canonical one-form that are written in terms of polynomials. This makes all the quantities appear in the central charge computation to be rational numbers.}
  Indeed, the same deformation has been already studied in \cite{Gadde:2013fma,Agarwal:2013uga,Agarwal:2014rua,Agarwal:2015vla}
  in the framework of class $\CS$ theories \cite{Gaiotto:2009we,Gaiotto:2009hg,Bah:2012dg}.
  We take the principal embeddings of all the $SU(N)^3$ flavor symmetry of $T_N$ theory. 
  Thus when $N=3$ this is not the principal embedding of the full flavor symmetry $E_6$. This deformation of the $T_N$ theory leads us to the $\CN=1$ SCFTs corresponding to the sphere with $0, 1, 2$ punctures. 
   
  The organization of this paper is as follows.
  In section \ref{sec:general}, 
  the general procedure of the deformation applicable to any $\CN=2$ SCFT with non-Abelian flavor symmetry is introduced.
  We give a formula of the 't Hooft anomaly coefficients, which is necessary for computing the IR R-symmetry.
  Sections \ref{sec:SU(2)} to \ref{sec:T_N} discuss examples of the flow by the principal embedding. 
  Then we conclude with some remarks in section \ref{sec:conclusion}.
  In addition we give a brief explanation on our convention of the $\CN=2$ $R$-charges and 't Hooft anomaly coefficients
  in appendix \ref{sec:app}.

\section{Deformation of $\CN=2$ SCFT with non-Abelian flavor symmetry}
\label{sec:general}
  In this section we consider the $\CN=1$ deformation procedure in a generic fashion.
  This is applied for any $\CN=2$ SCFT with non-Abelian flavor symmetry.
  
  Suppose we have an $\CN=2$ SCFT, $\CT$, with a non-Abelian flavor symmetry $F$.
  The $F$ could be a subgroup of the full flavor symmetry of $\CT$.
  The $R$-symmetry of $\CT$ is $SU(2)_R \times U(1)_{r}$.
  We denote the generators of the Cartan part of the $SU(2)_R$ and of $U(1)_{r}$ as $I_3$ and $r$ respectively.
  Due to the flavor symmetry, there exists the associated conserved current multiplet whose lowest component $\mu$ is the scalar
  with charge $(2I_3, r) = (2,0)$.
  
  We deform $\CT$ by adding an $\CN=1$ chiral multiplet $M$ transforming in the adjoint representation of $F$
  and the superpotential coupling
    \bea \label{eq:WmuM}
    W
     =     \Tr \mu M.
    \eea
  This superpotential breaks the supersymmetry to $\CN=1$.
  In the following we denote the $R$-symmetry of the theory as
    \bea
    2 I_3 
     =     J_+, ~~~
    r
     =     J_- ,
    \eea
  and sometimes set $R_0 = \half (J_+ + J_-)$. 
  The residual symmetry $\CF = \half (J_+ - J_-)$ is the global symmetry of the $\CN=1$ theory. 
  The $\CN=1$ $R$-charge in the $\CN=2$ algebra of the original theory $\CT$ is given by
  \be
   R_{\CN=1} = R_0 + \frac{1}{3} \CF = \frac{2}{3} J_+ + \frac{1}{3} J_- \ . 
  \ee
  The charges of $\mu$ and $M$ are $(J_+, J_-) = (2, 0)$ and $(J_+,J_-)=(0,2)$ respectively. 
  Even though the superpotential \eqref{eq:WmuM} makes the theory $\CN=1$ supersymmetric, 
  this term turns out to be irrelevant and the deformed theory in the IR simply decouples into the original $\CT$ and the free chiral multiplets $M$ 
  we added in the beginning. 
  
  A nontrivial $\CN=1$ fixed point can be produced by giving a nilpotent vev to $M$, 
  as in \cite{Heckman:2010qv,Gadde:2013fma,Agarwal:2013uga,Agarwal:2014rua, Agarwal:2015vla}.
  From the Jacobson-Morozov theorem, any nilpotent element of a semi-simple Lie algebra $\mathfrak{f}$ is given via embedding $\rho: \mathfrak{su}(2) \to \mathfrak{f}$ as $\rho(\s^+)$. Under the embedding, the adjoint representation of $\mathfrak{f}$ decomposes into
  \be
   \textrm{adj} \to \bigoplus_j V_j \otimes R_j \ , 
  \ee
  where $V_j$ is the spin-$j$ representation of $\mathfrak{su}(2)$ and $R_j$ is a representation under the commutant $\mathfrak{h}$ of $\mathfrak{f}$ under the embedding $\rho$. The commuting subgroup becomes the flavor symmetry of the theory after Higgsing. 
  
  When the flavor group is $F=SU(N)$, the vev is written in the block-diagonal Jordan form $\rho(\s^+) = \bigoplus_k  J_k^{\oplus n_k}$,
  where $J_k$ is the Jordan form of size $k$ and $n_k$ are integers.
  In other words this is specified by a partition of $N$:
    \bea
    N
     =     \sum_{k=1}^\ell k n_k.
    \eea
  Under the embedding, the adjoint representation decomposes into
    \bea
   {\rm adj} 
     &\rightarrow&    \bigoplus_{k<l} \bigoplus_{i=1}^{k} V_{\frac{l-k+2i-2}{2}} 
           \otimes \left( {\bf n}_{k} \otimes {\bf \bar{n}}_{l} 
           \oplus {\bf \bar{n}}_{k} \otimes {\bf n}_{l} \right)
           \oplus \bigoplus_{k=1}^{\ell} \bigoplus_{i=1}^{k} V_{i-1} 
           \otimes {\bf n}_{k} \otimes {\bf \bar{n}}_{k} - V_{0} \ .
           \label{adjointdecomposition}
    \eea
  The commutant $\mathfrak{h}$ of $\mathfrak{su}(N)$ under $\rho(\s^+)$ is given by $\mathfrak{s} [\prod_k \mathfrak{u}(n_k)]$,
  where $\mathfrak{s}$ means the overall traceless condition. 
    
    In this paper, we will mainly focus on the case of principal embedding $\rho$, which breaks the flavor symmetry $F$ completely upon Higgsing. 
    In this case, the adjoint representation of $\mathfrak{f}$ decomposes as
    \be
     \textrm{adj} \to \bigoplus_{i=1}^r V_{d_i - 1} \ , 
    \ee
  where $r = \textrm{rank}(\mathfrak{f})$ and $d_i$ are the degrees of Casimir invariants of $\mathfrak{f}$. 
  The degrees of invariants of the semi-classical Lie algebra are shown in table \ref{table:casimir}.
  The numbers $d_i - 1$ are also called the exponents of $\mathfrak{f}$. 
    \begin{table}
    \centering
      \begin{tabular}{|c||c|c}
      \hline
		$\mathfrak{f}$ & $d_i$ \\
		\hline \hline
		$\mathfrak{su}(n)$ & $2, 3, \ldots, n$ \\
		$\mathfrak{so}(2n+1)$ & $2, 4, 6, \ldots, 2n$ \\
		$\mathfrak{sp}(n)$ & $2, 4, 6, \ldots, 2n$ \\
		$\mathfrak{so}(2n)$ & $2, 4, 6, \ldots, 2n-2, n$ \\
		$\mathfrak{e}_6$ & $2, 5, 6, 8, 9, 12$ \\
		$\mathfrak{e}_7$ & $2, 6, 8, 10, 12, 14, 18$ \\
		$\mathfrak{e}_8$ & $2, 8, 12, 14, 18, 20, 24, 30$ \\
		$\mathfrak{f}_4$ & $2, 6, 8, 12$ \\
		$\mathfrak{g}_2$ & $2, 6$ \\
		\hline
	\end{tabular}
	\caption{Degrees of the Casimir invariants of the simple Lie algebras.} 
	\label{table:casimir}
    \end{table}
 
  Upon Higgsing via vev $\rho(\s^+)$, the superpotential term becomes
    \bea
    W
     =     \mu_{1,-1, 1} + \sum_{j,j_{3},f} {M}_{j,-j_{3},f} \mu_{j, j_{3},f},
           \label{vevsuperpote}
    \eea 
  where ${M}_{j, j_{3}, f}$ is the fluctuation of $M$ from the vev, 
  and $j$, $j_{3}$ and $f$ labels the spins, $\s_{3}$-eigenvalues 
  and the representations of the flavor symmetry $\mathfrak{h}$.
  Due to the first term of \eqref{vevsuperpote}, the $R$-symmetry gets shifted  
    \bea
     J_+ \rightarrow J_+, ~~~ 
     J_- \rightarrow J_- \  -2 \rho(\sigma^3) , 
    \label{chargeshift}
    \eea
   in order for the superpotential to have $(J_+, J_-)=(2, 2)$. 
  Furthermore the non-conservation of the flavor current $(D^2 J_F)_{j, j_3, f} = \delta W = \mu_{j, j_3 -1, f}$
  shows that the components of $\mu_{j, j_3, f}$ with $j_3 \neq j$ combine with the current and become non-BPS.
  The corresponding multiplets $M_{j, j_3, f}$ with $j_3 \neq -j$ thus decouple.
  The remaining multiplets $M_{j, -j, f}$ have charges $(J_+, J_-) = (0, 2 + 2 j)$, coupled to $\mu_{j,j,f}$. Therefore, we end up with the superpotential 
  \be 
   W = \sum_{j, f} M_{j, -j, f} \mu_{j, j, f} \ . 
  \ee
  When $\rho$ is the principal embedding, 
  we have $r$ chiral superfields $M_{j, -j}$ with $j = d_i - 1$ having charges $(J_+, J_-) = (0, 2d_i)$, $i=1, \ldots, r$. 

\paragraph{Chiral multiplets}
  The deformed theory has many $\CN=1$ chiral operators, in addition to $M_{j, -j, f}$.
  They come from the original theory $\CT$ which has Coulomb, Higgs and mixed branches.
  A Coulomb branch operator belongs to an $\CN=2$ short multiplet $\CE_{r(0, 0)}$ \cite{Dolan:2002zh} with $r=2\Delta$\footnote{Our convention is slightly different from the one in \cite{Dolan:2002zh} where $r$ charge was normalized to be equal to the scaling dimension for the Coulomb branch operators, so that $r_{\textrm{ours}} = 2 r_{\textrm{theirs}}$. }. 
  The components in the multiplets are
\be
 0_{(0, 0)}^r \to \left(\half \right)_{(0, \pm \half)}^{r-1} \to 0_{(0, \pm 1)}^{r-2}, 0_{(0, 0)}^{r-2}, 1_{(0, 0)}^{r-2} \to \left( \half\right)_{(0, \pm \half)}^{r-3} \to 0_{(0, 0)}^{r-4} 
\ee
  where $(I_3)_{(j_1, j_2)}^r$ stands for a component with spin $(j_1, j_2)$, $U(1)_r$ charge $r$ and $SU(2)_R$ charge $I_3$. The scaling dimension of the components are $\frac{r}{2}, \frac{r+1}{2}, \frac{r+2}{2}, \frac{r+3}{2}, \frac{r+4}{2}$ respectively. 
  
  This $\CN=2$ chiral multiplet can be decomposed into $\CN=1$ chiral multiplets. 
  (See appendix A of \cite{Gadde:2010en} for the detailed discussion on $\CN=1$ short multiplets.)
  In terms of their notation, the $\CN=2$ multiplet $\CE_{r(0, 0)}$ can be decomposed into
    \be
     \CE_{r(0, 0)} \to \CB_{\frac{r}{3} (0, 0)} \oplus \CB_{\frac{r+2}{3}(0, 0)} \oplus \CB_{\frac{r+1}{3} (0, \half)} \oplus \CB_{\frac{r+1}{3}(0, -\half)} \ , 
    \ee
  where the notation $\CB_{R_{\CN=1}(j_1, j_2)}$ stands for the short multiplet with $\CN=1$ $R$-charge $R_{\CN=1}$.
  The $\CN=1$ short multiplet $\CB_{R (j_1, j_2)}$ contains
    \be
    R^\Delta_{(j_1, j_2)} \to {(R-1)}^{\Delta+\half}_{(j_1, j_2 \pm \half)} \to (R-2)^{\Delta +1}_{(j_1, j_2)} \ , 
    \ee
  where $R^{\Delta}_{(j_1, j_2)}$ denotes operator with $R$-charge $R$, spin $(j_1, j_2)$ and dimension $\Delta$. 
  For completeness, let us write down $\CN=1$ chiral operators in $\CE_{r(0,0)}$ multiplet and their charges. 
    \be
    \begin{array}{|c||c|c|c|c|c|}
    \hline
    \CN=1 \textrm{ multiplet} & (j_1, j_2) & J_+ ~& J_- & \Delta_{{\rm UV}} = \frac{3}{2} {R_{\CN=1}} & R_{{\rm IR}} = \frac{1+\e}{2} J_+ + \frac{1-\e}{2} J_- \\
  \hline
 \CB_{\frac{r}{3}(0, 0)}& (0, 0)& 0 &  r & \frac{r}{2} & \frac{1-\e}{2} r \\
 \CB_{\frac{r+2}{3}(0, 0)}& (0, 0)&  2& r-2 & \frac{r}{2} + 1 & \frac{1-\e}{2} r +2 \e \\ 
 \CB_{\frac{r+1}{3} (0, \half)} & (0, \half) & 1 & r-1 & \frac{r+1}{2} & \frac{1-\e}{2} r + \e \\
 \CB_{\frac{r+1}{3} (0, -\half)} & (0, -\half) & 1 & r-1 & \frac{r+1}{2} & \frac{1-\e}{2} r + \e \\
 \hline
\end{array}
\ee
  Here the last column is the candidate $\CN=1$ $R$-charge after the deformation, as we will see later. 
  The UV $R$-charge $R_{\CN=1}$ is the same as $R_{{\rm IR}}$ with $\e = \frac{1}{3}$. 
   
  \paragraph{Central charges}
  The central charges, $a_{\CT}$ and $c_{\CT}$, and the flavor central charge $k_F$ of an $\CN=2$ SCFT $\CT$ can be written 
  in terms of the 't Hooft anomaly coefficients of the $R$-symmetries as follows \cite{Anselmi:1997am}:
    \begin{align}
    \begin{split}
    \tr J_+
    =     \tr J_+^3
    &=    0, \\
    \tr J_-
     =    \tr J_-^3
    &=   48(a_{\CT}-c_{\CT}), \\
    \tr J_+^2 J_- 
    &=   8 (2a_{\CT} - c_{\CT}),  \\
    \tr J_+ J_-^2
    &=    0,  \\
    \tr J_- T^a T^a
    &= - \frac{k_F}{2},
    \end{split}
    \label{N=2anom}
    \end{align}
  where $T_a$ are the generators of the flavor symmetry $F$.
  After the $\CN=1$ deformation, the $\tr J_-^3$ anomaly coefficient shifted due to \eqref{chargeshift} as 
    \bea
    \Tr J_-^3 \rightarrow \Tr J_-^3 + 12 \Tr J_- \rho(\sigma_3)^2, 
    \label{shiftJ_-}
    \eea
  while all the other anomalies remain fixed.
  The second term of \eqref{shiftJ_-} can be computed by remembering 
    \bea
    \Tr J_- \rho(\sigma_3)^2
     =     I_\rho \tr J_- T^a T^a
     =   - \frac{k_F}{2} I_\rho,
    \eea
  where the flavor index $a$ is not summed and $I_\rho$ is the embedding index. When $F = SU(N)$ where the embedding is given by the partition $N=\sum_k k n_k$, 
  the embedding index $I_\rho$ is given as 
    \bea
    I_{\rho} = \frac{1}{6} \sum_{k=1}^\ell k (k^2-1) n_k.
    \eea
  For the principal embedding associated with the partition $n_N=1$, this gives $I_\rho = \frac{N(N^2-1)}{6}$.
  In the case of the other semi-classical group, the embedding index for the principal embedding is given in table \ref{table:embidx}.

\begin{table}
\centering
\begin{tabular}{|c|ccccccccc|}
\hline 
 $\mathfrak{f}$ & $\mathfrak{su}(N)$ & $\mathfrak{so}(2N+1)$ & $\mathfrak{sp}(N)$ & $\mathfrak{so}(2N)$ & $\mathfrak{e}_6$ & $\mathfrak{e}_7$ & $\mathfrak{e}_8$ & $\mathfrak{f}_4$ & $\mathfrak{g}_2$ \\
 \hline 
 $I_{\textrm{pr}, \mathfrak{f}}$ & $\frac{N(N^2-1)}{6} $ & $\frac{N(N+1)(2N+1)}{3} $ & $\frac{N(4N^2-1)}{3}$ & $\frac{N(N-1)(2N-1)}{3}$ & $156$ & $399$ & $1240$ & $156$ & $28$ \\
 \hline
\end{tabular}
\caption{The embedding indices associated to the principal embeddings for all simple Lie algebras \cite{panyushev2009dynkin}.}
\label{table:embidx}
\end{table}

  By adding the contribution of the remaining chiral multiplet $M_{j,-j,f}$ one gets the anomalies of the deformed theory. 
  When $\rho$ is given by the principal embedding into $\mathfrak{f}$, the anomalies are given by
    \begin{align}
    \begin{split}
    \tr J_+  &=    \tr J_+^3     =   - r,   \\
    \tr J_- &=  48(a_{\CT}-c_{\CT}) + \sum_{i=1}^r (2d_i - 1), \\
    \tr J_-^3  &=   48(a_{\CT}-c_{\CT}) - 6k_F I_\rho + \sum_{i=1}^r (2d_i - 1)^3 , \\
    \tr J_+^2 J_-  &=    8 (2a_{\CT}-c_{\CT}) + \sum_{i=1}^r (2d_i - 1) ,          \\
    \tr J_+ J_-^2   &= - \sum_{i=1}^r (2d_i - 1)^2.
    \end{split}
    \label{eq:pDeformAnom}
    \end{align}
  For the $F=SU(N)$ case, they are
    \begin{align}
    \begin{split}
    \tr J_+
     &=    \tr J_+^3
     =   - N + 1,
               \\
    \tr J_-
    &=  48(a_{\CT}-c_{\CT}) + N^2 - 1,
              \\
    \tr J_-^3
    &=   48(a_{\CT}-c_{\CT}) - k_{SU(N)} N(N^2-1) + N^2(2N^2-1)- 1,
              \\
    \tr J_+^2 J_-
    &=    8 (2a_{\CT}-c_{\CT}) + N^2 - 1,
            \\
    \tr J_+ J_-^2
    &= -  \frac{N(4N^2-1)}{3}  + 1.
    \end{split}
    \label{totalanom}
    \end{align}

  By assuming that there is no accidental global symmetry in the IR, 
  the IR $U(1)_R$ symmetry is a combination of two $U(1)_{J_{\pm}}$ 
  which is determined by using $a$-maximization \cite{Intriligator:2003jj}. 
  Other $U(1)$ global symmetries in $\CT$ or those may come from the subgroup of $F$ cannot be mixed with $R$-symmetry. This is because $\tr R^2 F = 0$ for any global symmetry $F$ in an $\CN=2$ SCFT \cite{Kuzenko:1999pi}, and it also implies $\tr F = 0$ so that the flavor symmetry $F$ is ``baryonic".\footnote{We would like to thank Ken Intriligator for instructing this to us.} Baryonic symmetries cannot be mixed with the $R$-symmetry \cite{Intriligator:2003jj}.
  Therefore, we pick the trial $U(1)_R$ symmetry as
    \bea
    R
     =     \frac{1+\epsilon}{2} J_+ + \frac{1-\epsilon}{2} J_-
    \eea
  and compute the trial central charge $a(\epsilon) = \frac{3}{32}(3\tr R^3 - \tr R)$.
  Maximizing $a(\epsilon)$ gives a solution of $\epsilon$.
  However there is a caveat here: 
  one should check that all the dimensions of the operators, which is given by $\Delta(\CO_i) = \frac{3}{2} R(\CO_i)$, satisfy the unitarity bound.
  If it hits the bound, the operator becomes free and $U(1)$ symmetry under which only the free multiplet is charged appears. 
  One thus has to subtract the contribution of this operator from the trial central charge and re-maximize, as in \cite{Kutasov:2003iy}.
  This process has to be repeated until all the chiral operators satisfy the unitarity bound.


\section{Generalized Argyres-Douglas theories}
\label{sec:SU(2)}
  In this section we apply the general argument in the previous section to a family of SCFTs of Argyres-Douglas type 
  with an $SU(N)$ flavor symmetry.
  After reviewing the SCFTs, we consider the deformations of the so-called $(A_1, D_k)$ theory with an $SU(2)$ flavor symmetry, 
  and the $(I_{N,k}, F)$ theory with an $SU(N)$ flavor symmetry.
  Interestingly, for all these cases we will see the supersymmetry is enhanced to $\CN=2$ in the IR.  
 
\subsection{$\CN=2$ Argyres-Douglas theories}
\label{subsec:AD} 
  An $\CN=2$ SCFT with Coulomb branch operators with fractional dimensions is called as (a generalized) Argyres-Douglas theory. 
  The simplest example is the $H_0$ theory found in \cite{Argyres:1995jj,Argyres:1995xn} which has the single dimension-$\frac{6}{5}$
  operator.
  In the following, we will review the four classes of the Argyres-Douglas theories,
  collecting the results necessary for the computation in the subsequent subsection.
\paragraph{$(A_1, A_k)$ theory}
  This class of theories is obtained as the maximal conformal point on the moduli space of $\CN=2$ pure $SU(k+1)$ super Yang-Mills theory
  where mutually non-local massless particles appear.
  We assume that $k$ is greater than or equal to $2$.
  
  The central charges of the SCFT were given in \cite{Shapere:2008zf}:
    \bea
    a
     =     \frac{n(24n+19)}{24(2n+3)}, ~~~
    c
     =     \frac{n(6n+5)}{6(2n+3)}
            \label{acA1A2n-2}
    \eea
  for $k=2n$, and
    \bea
    a
     =    \frac{12n^2 +19n +2}{24(n+2)}, ~~~
    c
     =    \frac{3n^2 +5n +1}{6(n+2)}
           \label{acA1A2n-1}
    \eea
  for $k=2n+1$.
 
  From the fixed point, one can deform the theory by the Coulomb branch operator $\CO_i$ 
  (which is the lowest component of a $\CE_{r(0,0)}$ multiplet).
  The scaling dimensions of the operators are
    \bea \label{eq:A1A2nOps}
    \Delta(\CO_i)
     =     \frac{2(2n+3 - i)}{2n+3}, ~~~i=2,3,\ldots, n+1
    \eea
  for $k=2n$, and
    \bea
    \Delta(\CO_i)
     =     \frac{2n+4-i}{n+2}, ~~~i=2,3,\ldots,n+1
    \eea
  for $k=2n+1$.
  Their $R$-charges are $I_3(\CO_i)= 0$ and $r=2 \Delta(\CO_i)$.
  They span the Coulomb branch of the theory.   
  In the latter case we have a mass parameter with dimension $1$ associated to a $U(1)$ global symmetry.
  (For $k=3$ case it is enhanced to $SU(2)$ \cite{Argyres:2012fu}.)

\paragraph{$(A_1, D_k)$ theory}
  This class of theories is obtained as the maximal conformal point on the moduli space of $\CN=2$ $SU(k-1)$ gauge theory 
  with two fundamental hypermultiplets.
  The dimensions of the Coulomb branch operators $\CO_i$ are determined to be
    \bea
    \Delta(\CO_i)
     =     2 - \frac{2i}{k}, ~~~
    i
     =     1, 2, \ldots, [(k-1)/2]
    \eea
  where $[ \ldots ]$ is the integer part of $\ldots$.
  This class of theories has the $SU(2)$ flavor symmetry
  whose conserved current multiplet has a moment map operator $\mu$ as a lowest component, 
  and a corresponding mass parameter of dimension $1$.
  When $k=2n+2$, the flavor symmetry is $SU(2) \times U(1)$ as we can see that there is an additional dimension $1$ coupling. 
  Furthermore when $k=4$ this will enhance to $SU(3)$ \cite{Argyres:2012fu}.

  The central charges are given by \cite{Shapere:2008un}:
    \bea
    a
     =     \frac{n(8n+3)}{8(2n+1)}, ~~~
    c
     =     \frac{n}{2},
            \label{acA1D2n+1}
    \eea
  for $k=2n+1$, and
    \bea
    a
     =    \frac{n}{2} + \frac{1}{12}, ~~~
    c
     =    \frac{n}{2} + \frac{1}{6}
           \label{acA1D2n+2}
    \eea
  for $k=2n+2$.
  The flavor central charge of the $SU(2)$ symmetry is given by
    \bea
    k_{SU(2)}
     =     \frac{4(k-1)}{k}.
    \eea

\paragraph{$(A_{N-1}, A_{k-1})$ theory}
  Let us now see the generalization of these two classes.
  The generalization of the $(A_1, A_k)$ is the class $(A_{N-1}, A_{k-1})$ \cite{Cecotti:2010fi}.
  
  There are a large variety of theories depending on $k$.
  For simplicity, we will focus here on the special case where $k=Nm+N+1$.
  In this case, the dimensions of the Coulomb branch operators are 
    \bea
    \Delta(\CO_{i,j})
     =     \frac{Nj - (N-1)i}{Nm + 2N + 1},
     \label{dimAA}
    \eea
  where $i=2,3,\ldots,N$ and $j = m+2+i, \ldots, (m+2)i$.
  
  The central charges are computed in \cite{Xie:2013jc}
    \bea
    a
    &=&    \frac{(m+1)(N-1)N(4(m+1)N^2+4(m+3)N+3)}{48(N (m+2)+1)},
    \nonumber \\
    c
    &=&    \frac{(m+1)(N-1)N(N^2(m+1+N(m+3)+1))}{12(N (m+2)+1)}.
    \label{acAA}
    \eea
  See appendix \ref{subsec:conventionN=2} for a detailed computation.

\paragraph{$(I_{N,k}, F)$ theory}
  Then let us consider the so-called $(I_{N, k}, F)$ theory. 
  The scaling dimensions of the Coulomb branch operators $\CO_{i, j}$ are given by 
    \be
    \Delta(\CO_{i, j}) = \frac{k i - N j}{k+N} \ ,
    \label{dimINk}
    \ee
  where $i=2,3,\ldots, N$ and $j \geq - i$ such that the above dimension is greater than $1$. 
  For simplicity we will consider the case where $k = N m +1$.
  In this case the index $j$ runs from $-i$ to $i m$.
  This theory has a Higgs branch whose isometry is $SU(N)$.
  The mass parameters associated to it are $\CO_{i,-i}$ with $i=2,\ldots,N$.
  There is no other flavor symmetry,
  and thus this can be regarded as a natural generalization of the $(A_1, D_{2n+1})$ theory.
  
  The central charges of the theory with $k = N m +1$ are given by
    \bea
    a 
    &=&    \frac{\left(N^2-1\right) (k+N-1) (4 k+4 N-1)}{48 (k+N)},
               \nonumber \\
    c
    &=&    \frac{1}{12} (N+k-1)(N^2 - 1) \ , \quad k_{SU(N)} = \frac{2N(N+k-1)}{N+k} \ . 
    \eea
  See appendix \ref{subsec:conventionN=2} for the detail.

\subsection{Deformation of $(A_1, D_{2n+1})$ theory}
\label{subsec:D2n+1}
  Let us apply the deformation discussed in section \ref{sec:general} to the $(A_1, D_{2n+1})$ theory.
  The nilpotent vev specified by the partition of $2$, thus the only nontrivial vev is given by $n_1=0, n_2 =1$.
  
  The remaining component of $M$ is $M_{1,-1}$ with charges $(J_+,J_-)=(0,4)$.
  By using \eqref{totalanom}, the total anomaly coefficients of the deformed theory are
    \begin{align}
    \begin{split}
     & \tr J_+^3 = \tr J_+ = -1 \ , \\
     & \tr J_- =  \frac{3}{2n+1}, ~~~~
         \tr J_-^3 =  \frac{27}{2n+1} \ , \\
    & \tr J_+^2 J_-  = \frac{8n^2 + 8n+3}{2n+1} \ , ~~~~
        \tr J_+ J_-^2 = -9 \ . 
    \end{split}
    \end{align}
  By $a$-maximization, we obtain
    \be
    \e =\frac{-3 n^2+3 n+12+ \sqrt{36 n^4+156 n^3+241 n^2+136 n+16}}{3 \left(3 n^2+10 n+8\right)}. 
    \label{e1}
    \ee 

   As we have discussed in the previous section, an $\CN=2$ Coulomb branch multiplet can be decomposed in terms of $\CN=1$ multiplets as
    \be
    \CE_{r(0, 0)} \to \CB_{\frac{1-\e}{2} r(0, 0)} \oplus \CB_{(\frac{1-\e}{2}r + 2 \e)(0, 0)} \oplus \CB_{(\frac{1-\e}{2}r + \e) (0, \half)} 
    \oplus \CB_{(\frac{1-\e}{2}r + \e) (0, -\half)}  \ .
     \ee
  In the UV, before turning on the nilpotent deformation, $\e=\frac{1}{3}$ and $r = 2 \Delta(\CO_i)$, for the chiral multiplet containing $\CO_i$. 
  Now, with \eqref{e1}, we see that the multiplet $\CB_{\frac{1-\e}{2}r(0, 0)}$ which includes the operator $\CO_n$ 
  violates the unitarity bound.
  Thus, this indicates that the multiplet becomes free and decouple. 
  On the other hand, the rest of the $\CB$ short multiplets inside $\CE$ multiplets are above the unitarity bound, so they stay coupled. 
  
  Upon re-maximizing $a$, we obtain $\e = \frac{7+2n}{9+6n}$. 
  The Coulomb multiplets of the deformed $(A_1, D_{2n+1})$ theory in the IR are:
    \be \label{eq:EmultD2n}
    \CE_{\frac{4(2n+1-i)}{2n+1}(0, 0)} \to \CB_{\frac{4(2n+1-i)}{3(2n+3)}(0, 0)} \oplus \CB_{\frac{2(6n+9-2i)}{3(2n+3)}(0, 0)} 
    \oplus \CB_{\frac{10n+11-4i}{3(2n+3)}(0, \half)} \oplus \CB_{\frac{10n+11-4i}{3(2n+3)}(0, -\half)}  , ~~
    \ee
  for $i=1, 2, \cdots, n-1$, and 
    \be
    \CE_{\frac{4(n+1)}{2n+1} (0, 0)} \to \left( \CB_{\frac{4(n+1)}{3(2n+3)}(0, 0)} \right)_{\textrm{decoupled}} 
    \oplus \CB_{\frac{2(4n+9)}{3(2n+3)}(0, 0)} \oplus \CB_{\frac{6n+11}{6n+9}(0, \half)} \oplus \CB_{\frac{6n+11}{6n+9}(0, -\half)} .~~ 
    \ee
  Here the first $\CB$ multiplet in the parenthesis is the one that is decoupled along the RG flow and simply becomes the free chiral multiplet. 
  We also have $M_{1,-1}$ which is indeed the short multiplet $\CB_{\frac{4(2n+1)}{3(2n+3)}(0, 0)}$. 

  Now the central charges are
    \bea
    a
     =     \frac{n(24n+19)}{24(2n+3)}, ~~~
    c
     =     \frac{n(6n+5)}{6(2n+3)}.
    \eea
  One may notice that these are exactly the central charges of the $(A_1, A_{2n})$ theory.
  Indeed one could see this appearance of the $\CN=2$ SCFT in the IR by comparing the chiral operators as follows.
  The $(A_1, A_{2n})$ theory has $\CE$-type short multiplets that can be decomposed in terms of $\CN=1$ short multiplets: 
    \be \label{eq:EmultA2n}
    \CE_{\frac{4(2n+3-j)}{2n+3}(0, 0)} \to \CB_{\frac{4(2n+3-j)}{3(2n+3)}(0, 0)} \oplus \CB_{\frac{2(6n+9-2j)}{3(2n+3)}(0, 0)} 
    \oplus \CB_{\frac{10n+15-4j}{3(2n+3)}(0, \half)} \oplus \CB_{\frac{10n+15-4j}{3(2n+3)}(0, -\half)}  , ~~
    \ee
  where $j=2,3, \cdots, n+1$. 
  By comparing with \eqref{eq:EmultD2n}, one can see the matching of the chiral operators in the $\CE$-type chiral multiplets via 
    \begin{align}
	\left(\CO^{D_{2n+1}}_i \right)_{B_1} &\to \left( \CO^{A_{2n}}_{i+2}  \right)_{B_1} & (i=1, \cdots, n-1) \nn \\
	\left(\CO^{D_{2n+1}}_i \right)_{B_2} &\to \left( \CO^{A_{2n}}_{i}  \right)_{B_2} & (i=2, \cdots, n) \\
	\left(\CO^{D_{2n+1}}_i \right)_{F_{3, 4}} &\to \left( \CO^{A_{2n}}_{i+1}  \right)_{F_{3, 4}} & (i=1, \cdots, n) \nn
    \end{align}
  where $B_{1, 2}$ and $F_{3, 4}$ refers to the top components in the first and the latter two $\CN=1$ components in an $\CE$ multiplet. 
  We have one extra chiral multiplet $M_{1,-1}$, which is exactly the one corresponding to $\left( \CO^{A_{2n}}_2 \right)_{B_1}$. 

We seem to have one missing and one superfluous $\CB$ multiplet in this analysis to completely match with the Coulomb branch multiplets in the $(A_1, A_{2n})$ theory. We need 
\be \label{eq:A2nNeed}
 \left( \CO^{A_{2n}}_{n+1}\right)_{B_2} = \CB_{\frac{2(4n+7)}{3(2n+3)}(0, 0)} , 
\ee
and we have extra 
\be \label{eq:D2n1extra}
 \left(\CO^{D_{2n+1}}_1 \right)_{B_2} = \CB_{\frac{2(6n+7)}{3(2n+3)}(0, 0)} .
\ee
It is not so obvious from our analysis here whether we get the needed \eqref{eq:A2nNeed}, and whether \eqref{eq:D2n1extra} survives in the IR or not. From the superconformal index we compute in the section \ref{sec:SQCD}, we find that there is indeed the multiplet \eqref{eq:A2nNeed}, and the superfluous one \eqref{eq:D2n1extra} decouples. 

Other than subtleties regarding the short multiplets \eqref{eq:A2nNeed}, \eqref{eq:D2n1extra}, we have a nice match of $\CN=2$ chiral multiplets along the RG flow. One noticeable feature is that $\CN=1$ multiplets in each of $\CE$ multiplets in the IR comes from distinct $\CE$ multiplets in the UV. This shows that our deformation preserves only $\CN=1$ supersymmetry along the RG flow, but it enhances to $\CN=2$ in the IR.

\subsection{Deformation of $(A_1, D_{2n+2})$ theory} \label{sec:A1D2n2}
  Let us turn to the $(A_1, D_{2n+2})$ theory. 
  The anomaly coefficients of the deformed $(A_1, D_{2n+2})$ theory after the Higgsing are
   \begin{align}
   \begin{split}  
 &   \Tr J_+ = \Tr J_+^3 =    -1 , \\
 &  \Tr J_-  = - 1, ~~~~
    \Tr J_-^3     =     \frac{-n+11}{n+1}, \\
 &  \Tr J_+^2 J_- =   4n + 3, ~~~~
    \Tr J_+ J_-^2 =  - 9.
   \end{split}
    \end{align}
  The trial central charge is maximized at $\epsilon = \frac{n+4}{3(n+2)}$, which gives
    $a
     =     \frac{8n^2 + 13n + 2}{16n+32}$ and 
    $c
     =     \frac{4n^2 + 7n + 2}{8n + 16}$.
     
  The $\CE$-multiplets of the $(A_1, D_{2n+2})$ theory in the UV decomposes into $\CN=1$ multiplets in the IR as
    \be
	\CE_{\frac{4(2n+2-i)}{2n+2}(0, 0)}  \to \CB_{\frac{2(2n+2-i)}{3(n+2)}(0, 0)} \oplus \CB_{\frac{2(3n+6-i)}{3(n+2)}(0, 0)} \oplus \CB_{\frac{5n+8-2i}{3 (n+2)}(0,  \half)} \oplus \CB_{\frac{5n+8-2i}{3 (n+2)}(0, -\half)}  ,
    \ee
  where $i=1, 2, \cdots n$. When $i=n$, the first $\CB$ multiplet have $R$-charge $\frac{2}{3}$ and becomes free. 
  Thus this operator decouples, leaving dimension one mass parameter.
  Upon subtracting the contribution of the decoupled chiral multiplet, we get
    \bea
    a
     =     \frac{12n^2 + 19n + 2}{24n+48}, ~~~
    c
     =     \frac{3n^2 + 5n + 1}{6n + 12},
    \eea
  which are exactly the same as those of the $(A_1, A_{2n+1})$ theory.
  
  The $\CE$-multiplets in the $(A_1, A_{2n+1})$ theory are given by
    \be
	\CE_{\frac{2(2n+3-j)}{n+2}(0, 0)} \to \CB_{\frac{2(2n+3-j)}{3(n+2)}(0, 0)} \oplus \CB_{\frac{2(3n+5-j)}{3(n+2)}(0, 0)} \oplus \CB_{\frac{5n+8-2j}{3(n+2)}(0, \half)} \oplus \CB_{\frac{5n+8-2j}{3(n+2)}(0, -\half)} , 
    \ee
  with $j=1, 2, \cdots, n$. 
  We see that the above bosonic $\CB$-multiplets inside the $\CE$-multiplets match upon $i = j-1$ for the first, and $i=j+1$ for the second. 
  The Fermionic ones match with $i=j$. 
  The first missing bosonic $\CB$-multiplets ($j=1$) comes from the $M_{1,-1}$ field with the $R$-charge $\frac{4(n+1)}{3(n+2)}$ in the IR. 

  Therefore we have the matching of the spectrum of Coulomb branch operators 
  except for the superfluous one $\CB_{\frac{2(3n+5)}{3(n+2)}(0, 0)}$ and the missing one $\CB_{\frac{2(2n+5)}{3(n+2)}(0, 0)}$. 
  It is not clear to us from here, how the superfluous operator decouple and the missing one appears along the RG flow. 
  From the superconformal index we compute in section \ref{sec:SQCD}, 
  we see that $\CB_{\frac{2(3n+5)}{3(n+2)}(0, 0)}$ multiplet is removed along the flow and the $\CB_{\frac{2(2n+5)}{3(n+2)}(0, 0)}$ multiplet indeed appears.

\subsection{Deformation of $(I_{N,k}, F)$ theory}
\label{subsec:SU(N)}
  Now we turn to a case with an $SU(N)$ flavor symmetry. 
  As we review in section \ref{subsec:AD}, we focus here on the case with $k=Nm+1$.
  Let us consider the deformation corresponding to the principal embedding: the partition specified by $n_N=1$. 
  Due to the Higgsing, the remaining components of $M$ are the ones $M_{j, -j}$ with $j=1, \ldots, N-1$ 
  where the charges are given by $(J_+, J_-) = (0, 2j+2)$. 
  Thus from  \eqref{totalanom}, the total anomalies are obtained as
     \begin{align}
    \begin{split}
 & \tr J_+ = \tr J_+^3 = -N+1 \ , \\
 & \tr J_- = \frac{N^2 - 1}{N+k} \ ,~~~~
   \tr J_-^3 =  \frac{2 N^4-N^2-1}{k+N} \ , \\
 & \tr J_+^2 J_- =  \frac{\left(N^2-1\right) \left(2 k^2+4 k N+2 N^2+1\right)}{3 (k+N)} \ , \\
 & \tr J_+ J_-^2 = \frac{1}{3} \left(-4 N^3+N+3\right) \ .
\end{split}
\end{align}
  By performing $a$-maximization with the above anomaly coefficients we find the $N-1$ Coulomb branch operators $\CO_{i,j}$, with $(i,j)=(i,(i-1)m-1)$, hit the unitarity bound.
  By subtracting these contributions as in the previous section, and $a$-maximizing again, we obtain
    \bea
    \epsilon
     =     \frac{N m + 4N + 1}{3(N m + 2 N +1)}
    \eea
  and the central charges
    \bea
    a
    &=&    \frac{(m+1)(N-1)N(4(m+1)N^2+4(m+3)N+3)}{48(N (m+2)+1)},
    \nonumber \\
    c
    &=&    \frac{(m+1)(N-1)N(N^2(m+1+N(m+3)+1))}{12(N (m+2)+1)}.
    \eea
  The dimensions of the remaining operators are now given by
    \bea
    [\hat{u}_{i,j}]
     =     \frac{(N m+1)i- Nj}{N (m+2)+1}
     \label{dimnew}
    \eea
  where $i=2,3,\ldots,N$ and $j=-i, -i+1,\ldots,(i-1)m-2$.
  The operators $(i,j) = (i,-i)$ come from $M_{i-1,-(i-1)}$.  
  
  We note that these dimensions and the central charges agree with \eqref{dimAA} and \eqref{acAA} in section \ref{subsec:AD}.
  Thus we conclude that by the deforming the $(I_{N, Nm+1}, F)$ theory corresponding to the principal nilpotent element, one gets the $(A_{N-1}, A_{Nm+N})$ theory in the IR.

\section{The rank-one SCFTs}
\label{sec:rank1}
  In this section, we consider the $\CN=1$ deformations of rank-one SCFTs with non-Abelian flavor symmetries.
  By rank we mean the complex dimension of the Coulomb branch. 
  A classification of the $\CN=2$ rank-one SCFTs has been performed in \cite{Argyres:2015ffa,Argyres:2015gha}
  from the perspective of the Coulomb branch geometry, which is restricted to be the singularities of Kodaira type.
  The possible (relevant) deformations from these geometries classify the $\CN=2$ rank-one theories.
  
  The first series of the theories which we will discuss in subsection \ref{subsec:H0} is specified 
  by the ``maximal" deformations of the Kodaira singularities.
  This leads to the SCFTs which we call as $H_{0,1,2}$, $D_4$, and $E_{6,7,8}$.
  These were found originally in \cite{Seiberg:1994aj, Argyres:1995jj,Argyres:1995xn,Minahan:1996fg,Minahan:1996cj}. 
  We will see that for all of these theories, the $\CN=1$ deformations associated to the principal embedding lead to the $H_0$ theory in the IR with some decoupled chiral multiplets. 
  In subsection \ref{subsec:other}, we deal with some of the other SCFTs associated to the other deformations of the Coulomb branch geometry.

\subsection{Flows to $H_0$ $\CN=2$ SCFT}
\label{subsec:H0}
  Let us consider rank-one SCFTs 
  $H_0, H_1, H_2, D_4, E_6, E_7, E_8$,
  where the $H_0$, $H_1$, $H_2$ theories are the same as $(A_1, A_2)$, $(A_1, A_3)=(A_1, D_3)$ 
  and $(A_1, D_4)$ Argyres-Douglas theories respectively. 
  We summarize the central charges and dimensions of the Coulomb branch operators in the table \ref{table:rank1}. 
  They also have a simple realization as world-volume theories on a D3-brane in F-theory singularities 
  \cite{Banks:1996nj,Dasgupta:1996ij,Sen:1996vd}.   
  
  We have already found that the $\CN=1$ principal deformation of the $H_1$ theory (which is $(A_1, D_3)$) leads to the $H_0$ theory
  in section \ref{sec:SU(2)}. 
  We examine the remaining cases in this subsection. 

\begin{table}[t] 
\renewcommand{\arraystretch}{1.2} 
\centering
\begin{tabular}{|c|ccccccc|}
\hline
 $G$ & $H_0$ & $H_1$ & $H_2$ & $D_4$ & $E_6$ & $E_7$ & $E_8$ \\
 \hline 
 $k_G$ &$\cdot$ &  $\frac{8}{3}$ & $3$ & $4$ & $6$ & $8$ & $12$ \\
 $a$ & $\frac{43}{120}$ & $\frac{11}{24}$ & $\frac{7}{12}$ & $\frac{23}{24}$ & $\frac{41}{24}$ & $\frac{59}{24}$ & $\frac{95}{24}$ \\
 $c$ & $\frac{11}{30}$ & $\frac{1}{2}$ & $\frac{2}{3}$ & $\frac{7}{6}$ & $\frac{13}{6}$ & $\frac{19}{6}$ & $\frac{31}{6}$  \\
 $\Delta(u)$ & $\frac{6}{5}$ & $\frac{4}{3}$ & $\frac{3}{2}$ & $2$ & $3$ & $4$ & $6$  \\
 \hline
\end{tabular}
\caption{The central charges and the dimensions of the Coulomb branch operators of the rank-one SCFTs \cite{Argyres:2007cn, Aharony:2007dj,Shapere:2008zf, Argyres:2007tq}.}
\label{table:rank1}
\end{table} 

\subsubsection*{$H_2$ theory}
  We considered the deformation of $H_2 = (A_1, D_4)$ theory in section \ref{sec:A1D2n2}. 
  There we only considered a deformation breaking the $SU(2)$ ($\subset SU(3)$) flavor symmetry leaving $U(1)$ symmetry. 
  We have observed that in this case the $U(1)$ symmetry is actually enhanced to $SU(2)$ symmetry. 
  Here, let us consider breaking the entire $SU(3)$ flavor symmetry. 

  Under the principal embedding, the adjoint representation of $SU(3)$ decomposes into $ 8 \to  V_1 \oplus V_2$, 
  where $V_j$ is the spin-$j$ irreducible representation of $SU(2)$. 
  We are now left with $M_{j, -j}$ with $j=1, 2$ with $(J_+, J_-) = (0, 4), (0, 6)$. 
  The anomalies after the deformation are given by
    \bea
    \tr J_+
    &=&    \tr J_+^3 
     =   -2 ,   \nonumber \\
    \tr J_-
    &=&    4, ~~~~~~~
    \tr J_-^3
     =      76,   
               \\
    \tr J_+^2 J_- 
    &=&    12, ~~~~~
    \tr J_+ J_-^2
     =   - 34. \nonumber
    \eea
  From here, we obtain the trial $a$-function as $a(\e) = \frac{3}{32} \left(81 \epsilon ^3-108 \epsilon ^2+33 \epsilon -2\right)$, 
  which upon $a$-maximization, we get $ \e = \frac{1}{9} \left(\sqrt{5}+4\right) \simeq 0.692896 $. 
  This makes the $M_{1, -1}$ and the Coulomb branch operator (having $(J_+, J_-)=(0, 3)$) to violate the unitarity bound. 
  Therefore they have to be decoupled. 
  After decoupling, the anomalies are
    \bea
    \tr J_+
    &=&    \tr J_+^3 
     =   -2 ,   \nonumber \\
    \tr J_-
    &=&    -1, ~~~~~
    \tr J_-^3
     =      41,   
               \\
    \tr J_+^2 J_- 
    &=&    7, ~~~~~~~
    \tr J_+ J_-^2
     =   - 21, \nonumber
    \eea
  which gives the trial $a$-function to be $a(\e) = - \frac{3}{256}  \left(375 \epsilon ^3-495 \epsilon ^2+121 \epsilon -1\right)$. 
  Now, we obtain $\e = \frac{11}{15}$, from which the central charges are calculated as
    \be
    a = \frac{43}{120} \ , \qquad c = \frac{11}{30}.
         \label{centralH0} 
    \ee
  These are exactly the same values as those of the $H_0$ (or $(A_1, A_2)$) theory. 
  We also find that the operator $M_{2, -2}$ has the conformal dimension $\Delta = \frac{6}{5}$, 
  which is the same as that of the Coulomb branch operator of the $H_0$ theory. 
  Therefore we have found an RG flow that takes the $H_2$ theory (with chiral multiplets) to $H_0$ (with some free chiral multiplets).

\subsubsection*{$D_4$ theory} \label{subsec:D4}
  Let us consider the $D_4$ theory, which is the $\CN=2$ SCFT realized by $SU(2)$ theory with 4 fundamental hypermultiplets. 
  We couple $28$ chiral multiplet $M$ with the $SO(8)$ moment map operator $\mu$ via $W=\tr M \mu$. 
  We give a nilpotent vev corresponding to the principal embedding of $SU(2)$ into $SO(8)$. 
  Under the principal embedding, the adjoint representation of $SO(8)$ decomposes into
    \be
    28 \to V_1 \oplus V_3 \oplus V_5 \oplus V_3.
    \ee
  Upon giving the vev to $M$, we are left with $M_{j, -j}$ with $j=1, 3, 5, 3$ with $(J_+, J_-) =(0, 2 + 2j)$.
  
  The anomalies after the deformation are given by
    \bea
    \tr J_+
    &=&    \tr J_+^3 
     =   -4 ,   \nonumber \\
    \tr J_-
    &=&    18, ~~~~~
    \tr J_-^3
     =      1362,   
               \\
    \tr J_+^2 J_- 
    &=&    34, ~~~~~
    \tr J_+ J_-^2
     =   - 228, \nonumber
    \eea
  from which we get the trial $a$-function as $a(\e) = - \frac{3}{32}  \left(807 \epsilon ^3-1746 \epsilon ^2+1231 \epsilon -284\right)$. 
  Upon $a$-maximization, we get $ \e = \frac{1}{807} \left(582+\sqrt{7585}\right) \simeq 0.82911$. 
  This makes the Coulomb branch operator (that has $(J_+, J_-)=(0, 4)$) and $M_{1, -1}$ to violate the unitarity bound 
  so that they become free along the RG flow and get decoupled. 
  
  The computation after was reported in \cite{Maruyoshi:2016tqk}: we redo the $a$-maximization twice due to the unitarity violating operators, 
  and the resulting central charges are the same as those of the $H_0$ theory \eqref{centralH0}.
  We also find that the operator $M_{5, -5}$ has the conformal dimension $\Delta = \frac{6}{5}$. 
  Therefore we have found an RG flow that takes the $D_4$ theory to $H_0$ (with some free chiral multiplets). 

\paragraph{Lagrangian after the nilpotent Higgsing}
  Since the $D_4$ theory has a Lagrangian description, 
  we can write down the matter content after integrating out massive modes from the Higgsing. 
  The procedure is essentially the same as the one considered in \cite{Agarwal:2014rua,Agarwal:2015vla}. 
  
  Before the Higgsing, the matter content is simply given by that of $\CN=2$ $SU(2)$ gauge theory 
  with eight fundamental half-hypermultiplets $q$ 
  and a chiral multiplet $M$ transforming in the adjoint representation of the flavor group $SO(8)$. 
  The charges of the superfields are given in table \ref{tab:D4}. 
  The superpotential is simply given by
    \be
    W = \tr \phi \tilde{\mu} + \tr M \mu , 
    \ee 
  with $\mu^{ij} = \e^{\a \b} q_\a^i q_\b^j $ and $\tilde{\mu}_{\a \b} = \delta_{ij} q^i_\a q^j _\b$ 
  where $\a, \b$ are gauge indices and $i, j$ are the flavor indices, and $\phi$ is the adjoint chiral superfield in the $\CN=2$ vector multiplet.
  
  \begin{table}[t]
  \begin{center}
  \begin{tabular}{cc}
  \begin{minipage}{0.5\hsize}
  \begin{center}
  \begin{tabular}{|c|ccc|}
  \hline
	 & $SU(2)$ & $SO(8)$ & $(J_+, J_-)$ \\
	 \hline 
	 $q$ & $\square$ & $\square$ & (1, 0) \\
	 $\phi$ & adj & 1 & (0, 2) \\
	 $M$ & 1 & adj & (0, 2) \\
	 \hline
	\end{tabular}
    \caption{Charges of the fields of the $SU(2)$ gauge theory with eight half-hypermultiplets.}
    \label{tab:D4}
   \end{center}
  \end{minipage}
  \begin{minipage}{0.5\hsize}
  \begin{center}
  \begin{tabular}{|c|cc|}
  \hline
  & $SU(2)$ & $(J_+, J_-)$ \\
  \hline 
  $q$ & $\square$ & (1, 0) \\
  $q'$ & $\square$ & (1, -6) \\
  $\phi$ & adj & (0, 2) \\
  $M_{1}$ & 1 & (0, 4) \\
  $M_{3}$ & 1 & (0, 8) \\
  $M_{5}$ & 1 & (0, 12) \\
  $M'_{3}$ & 1 & (0, 8) \\
  \hline
  \end{tabular}
  \caption{Charges of the fields after the Higgsing due to the nilpotent vev.}
  \label{tab:D42}
  \end{center}
  \end{minipage}
  \end{tabular}
  \end{center}
  \end{table}

  After the Higgsing, the $(J_+, J_-)$ charges are shifted according to \eqref{chargeshift}. 
  As above we are left with 4 components $M_{j, -j}$ with $j=1, 3, 5, 3$. 
  In order to see the remaining quarks, note that the fundamental representation of $SO(8)$ decomposes $8 \to 7 \oplus 1 =  V_3 \oplus V_0$ 
  under the principal embedding. 
  Therefore, we are left with 2 doublets of $SU(2)$, with charges $(1, 0)$ and $(1, -6)$. 
  To summarize, we get the matter content in table \ref{tab:D42}.
  The superpotential is given by
    \be
    W = \phi q q +  \sum_{j=1, 3, 5, 3'} \mu_{j} M_{j, -j} \ , 
    \ee
  where $j=3'$ means $\mu_3'$ and $M'_{3, -3}$. The $\mu_j$ operators are given by appropriate combination of $q, q'$ and $\phi$ 
  to have the appropriate charges $(2, -2j)$. It leaves us with the unique choice
    \be
    \mu_1 = \phi q q' , \quad \mu_3 = q q', \quad \mu_5 = \phi q' q' , \quad \mu_3' = \phi^3  q' q'  \ , 
    \ee
  where we omitted the indices. 
  This gauge theory preserves $U(1)_\CF \times U(1)_R$ global symmetry, which gets enhanced to $SU(2)_R \times U(1)_r$ in the IR. 
  This is the $\CN=1$ Lagrangian gauge theory flows to the ``non-Lagrangian" Argyres-Douglas theory $H_0$. 
  This result has been reported in \cite{Maruyoshi:2016tqk}, 
  where we computed the full superconformal index of the $H_0$ theory using this ``Lagrangian" description. 

\subsubsection*{$E_6$ theory}
  Let us consider the deformation of the $E_6$ SCFT. 
  Let us add $78$ chiral multiplets $M$ to couple with the moment map operator of the $E_6$ flavor symmetry via $W = \tr M \mu$.
  Under the principal embedding, the adjoint representation of $E_6$ decomposes into the $SU(2)$ representation as 
    \be
    78 \to V_1 \oplus V_4 \oplus V_5 \oplus V_7 \oplus V_8 \oplus V_{11} .
   \ee
  We give the vev to $M$ according to this embedding, which leaves all the $E_6$ symmetry to be broken, 
  and six components $M_{j, -j}$ with $j=1, 4, 5, 7, 8, 11$ which have $(J_+, J_-) = (0, 2 + 2j)$.
  
  By recalling that the embedding index for the principal embedding is $I_{E_6} = 156$ (see the table \ref{table:embidx}),
  the anomaly coefficients are given by
    \bea
    \tr J_+
    &=&    \tr J_+^3 
     =   - 6,   \nonumber \\
    \tr J_-
    &=&    56, ~~~~~
    \tr J_-^3
     =      16904,   
               \\
    \tr J_+^2 J_- 
    &=&    88, ~~~~~
    \tr J_+ J_-^2
     =   - 1254. \nonumber
    \eea
  This leads to the trial function $a(\e) = -\frac{3}{32} \left(7851 \epsilon ^3-20322 \epsilon ^2+17483 \epsilon -5000\right)$, 
  which upon $a$-maximization gives  $ \e = \frac{6774+\sqrt{134065}}{7851} \simeq 0.909457$. 
  This makes $M_{j, -j}$ with $j=1, 4, 5$ and the Coulomb branch operator to have $R$-charges below the unitarity bound.  

  Let us decouple these unitarity violating operators. 
  Then we get the 't Hooft anomaly coefficients 
    \bea
    \tr J_+
    &=&    \tr J_+^3 
     =   - 2,   \nonumber \\
    \tr J_-
    &=&    28, ~~~~~
    \tr J_-^3
     =      14692,   
               \\
    \tr J_+^2 J_- 
    &=&    60, ~~~~~
    \tr J_+ J_-^2
     =   - 1018. \nonumber
    \eea
  from which we obtain the $a$-function to be $a(\e) =  - \frac{3}{32} \left(6723 \epsilon ^3-17604 \epsilon ^2+15303 \epsilon -4418\right)$. 
  From $a$-maximization, we get $ \e = \frac{1}{747} \left(652+\sqrt{1721}\right) \simeq 0.92836$. 
  This makes $M_{j, -j}$ with $j=7, 8$ to violate the unitarity bound. 

  Decoupling these operators as well, finally we get
    \bea
    \tr J_+
    &=&    \tr J_+^3 
     =   0,   \nonumber \\
    \tr J_-
    &=&  -4, ~~~~~
    \tr J_-^3
     =      6404,   
               \\
    \tr J_+^2 J_- 
    &=&    28, ~~~~~
    \tr J_+ J_-^2
     =   - 504. \nonumber
    \eea
  and $a(\e) = -\frac{3}{8}  \left(750 \epsilon ^3-1935 \epsilon ^2+1652 \epsilon -467\right)$.
  Upon $a$-maximization, we get $\e = \frac{14}{15}$. 
  This gives us the central charges of the $H_0$ theory (once we throw away 6 decoupled free chiral multiplets). 
  Note that the conformal dimension of the $M_{11, -11}$ operator is $\Delta = \frac{6}{5}$, 
  which is the same as that of the Coulomb branch operator of the $H_0$ theory. 

  There is an $\CN=1$ gauge theory flows to this $\CN=2$ $E_6$ SCFT in the infrared \cite{Gadde:2015xta}, which provides a physical interpretation of the computation of the superconformal index done in \cite{Gadde:2010te}. 
  Combined with our result in this section, we have an alternative UV or a ``dual" description of the $H_0$ theory with different matter content and gauge group.

\subsubsection*{$E_7$ theory}
  Let us consider the $\CN=2$ SCFT with the $E_7$ global symmetry. 
  Under the principal embedding, the adjoint of the $E_7$ decomposes into
    \be
   133 \to V_1 \oplus V_5 \oplus V_7 \oplus V_9 \oplus V_{11} \oplus V_{13} \oplus V_{17}
    \ee
  From here, we obtain the anomaly coefficients after Higgsing to be
    \bea
    \tr J_+
    &=&    \tr J_+^3 
     =   -7,   \nonumber \\
    \tr J_-
    &=&   99, ~~~~~
    \tr J_-^3
     =      67131,   
               \\
    \tr J_+^2 J_- 
    &=&    147, ~~~~~
    \tr J_+ J_-^2
     =   -3199. \nonumber
    \eea
  The repeated $a$-maximization as the $E_6$ case makes $M_{j, -j}$ with $j=1, 5, 7, 9, 11, 13$ and the Coulomb branch operator 
  (having $(J_+, J_-)=(0, 8)$) to violate the unitarity bound. 
  After all, we obtain the same central charges as those of the $H_0$ theory. 
  The $M_{13, -13}$ has the scaling dimension $\frac{6}{5}$. 
  Therefore we obtain $H_0$ theory at the end of the RG flow. 

\subsubsection*{$E_8$ theory}
  Let us consider the $\CN=2$ SCFT with $E_8$ the global symmetry. 
  The adjoint representation of $E_8$ decomposes under the principal embedding as 
    \begin{align} \begin{split}
    248 \to V_1 \oplus V_7 \oplus V_{11} \oplus V_{13} \oplus V_{17} \oplus V_{19} \oplus V_{23} \oplus V_{29} \ . 
    \end{split} \end{align}
  The anomalies after Higgsing are given by 
    \bea
    \tr J_+
    &=&    \tr J_+^3 
     =   -8,   \nonumber \\
    \tr J_-
    &=&   190, ~~~~~
    \tr J_-^3
     =      357310,   
               \\
    \tr J_+^2 J_- 
    &=&    270, ~~~~~
    \tr J_+ J_-^2
     =   -9928. \nonumber
    \eea
  We repeat the same procedure as before multiple times to find that the only $M_{j, -j}$ operator that remains coupled is the one with $j=29$. 
  Then this again leads to the central charges $a=\frac{43}{120}$ and $c=\frac{11}{30}$ and $\Delta(M_{29, -29}) = \frac{6}{5}$. 
  Therefore we end up with the $H_0$ theory as the previous examples.

\subsection{Other rank-one SCFTs}
\label{subsec:other}
  Let us consider other rank-one SCFTs with non-Abelian flavor symmetries found in \cite{Argyres:2007tq, Argyres:2015ffa,Argyres:2015gha,Chacaltana:2016shw, Argyres:2016xua}. 
  We consider here some of the SCFTs obtained by non-maximal deformation of the Coulomb branch geometry of 
  Kodaira type $IV^*$, $III^*$ and $II^*$, and $\CN=4$ $SU(2)$ super Yang-Mills theory, which are listed in the table 1 in \cite{Argyres:2016xua}.
  While these are specified by the type of geometries, we refer to these in terms of their flavor symmetries. 
  From the $IV^*$ geometry we have an SCFT with $Sp(2) \times U(1)$ flavor symmetry while the maximal deformation gives the $E_6$ theory.
  There are SCFTs with $SU(2) \times U(1)$ and $Sp(3) \times SU(2)$ flavor symmetries from the $III^*$ type,
  and the ones with $SU(3)$, $SU(4)$ and $Sp(5)$ from the $II^*$ type.
  
  We will consider these theories in order and find that they in general flow to $\CN=1$ SCFTs.
  For the $Sp(2) \times U(1)$ and $SU(3)$ theories we get rational central charges. 
  We are not sure whether there is the enhancement of the supersymmetry in these cases,
  as we could not figure out these from the known central charges of the $\CN=2$ SCFTs.

\paragraph{$\CN=4$ SU(2) super Yang-Mills theory}
  The $\CN=4$ super Yang-Mills theory is regarded as
  the $\CN=2$ gauge theory with the hypermultiplet transforming in the adjoint representation of the gauge group.
  When the gauge group is $SU(2)$ this is rank-one SCFT, with Coulomb branch operator of dimension $2$.
  Also in this case the flavor symmetry is $SU(2)$, thus we can perform our $\CN=1$ deformation.
  The central charges are easy to obtain as
    \bea
    a
     =     c
     =     \frac{3}{4}, ~~~~
    k_{SU(2)}
     =     4.
    \eea
    
  Let us consider the $\CN=1$ deformation associated with the principal embedding of $SU(2)$ flavor symmetry.
  After the deformation, the remaining component of $M$ is only $M_{1,-1}$ with charges $(J_+, J_-) = (0,4)$.
  The 't Hooft anomaly coefficients
    \bea
    \tr J_+
    &=&    \tr J_+^3 
     =     - 1,   \nonumber \\
    \tr J_-
    &=&   3, ~~~~~
    \tr J_-^3
     =      9,   
               \\
    \tr J_+^2 J_- 
    &=&    9, ~~~~~
    \tr J_+ J_-^2
     =   - 9. \nonumber
    \eea  
  The $a$-maximization gives $\e = \frac{3+ \sqrt{97}}{24}$,
  and there is no operator which violates the unitarity bound.
  The IR theory is an $\CN=1$ SCFT with irrational central charges $a \simeq 0.6362$ and $c \simeq 0.6406$.
  Both $u$ and $M_{1,-1}$ have the same dimension $\Delta(u) = \Delta(M_{1,-1}) \simeq 1.3939$.

\paragraph{$Sp(2) \times U(1)$ theory} 
  The central charges are 
    \bea
    a
     =     \frac{17}{12}, ~~~~
    c
     =     \frac{19}{12}, ~~~~
    k_{Sp(2)}
     =     4.
    \eea
  The Coulomb branch operator has dimension $3$.
  
  In this case the $\CN=1$ deformation leaves us $M_{j,-j}$ with charge $(0,2+2j)$ where $j=2,4$.
  The 't Hooft anomaly coefficients are given by
    \bea
    \tr J_+
    &=&    \tr J_+^3 
     =     - 2,   \nonumber \\
    \tr J_-
    &=&   2, ~~~~~
    \tr J_-^3
     =      122,   
               \\
    \tr J_+^2 J_- 
    &=&    20, ~~~~~
    \tr J_+ J_-^2
     =   - 58. \nonumber
    \eea  
  The $a$-maximization gives $\e = \frac{237+ 2\sqrt{5137}}{537}$, which means that $M_{2,-2}$ violates the unitarity bound. 
  After subtracting this contribution and $a$-maximizing again, we obtain $\e = \frac{5}{7}$ and the central charges
    \bea
    a
      =     \frac{87}{112}, ~~~
    c
      =     \frac{47}{56}.
    \eea
  The scaling dimensions of the chiral operators are $\Delta(M_{4,-4}) = \frac{12}{7}$ and $\Delta(u) = \frac{9}{7}$.
  While it has rational central charges, we are not aware of whether the SCFT has $\CN=2$ supersymmetry or not.

\paragraph{$SU(2) \times U(1)$ theory}
  The central charges are
    \bea
    a
     =     \frac{15}{8}, ~~~~
    c
     =     2, ~~~~
    k_{SU(2)}
     =     10.
    \eea
  The Coulomb branch operator $u$ has dimension $4$.
  
  After the deformation we have $M_{1,-1}$ with charge $(0,4)$.
  The 't Hooft anomaly coefficients
    \bea
    \tr J_+
    &=&    \tr J_+^3 
     =     - 1,   \nonumber \\
    \tr J_-
    &=&  -3, ~~~~~
    \tr J_-^3
     =     - 39,   
               \\
    \tr J_+^2 J_- 
    &=&    17, ~~~~~
    \tr J_+ J_-^2
     =   - 9. \nonumber
    \eea  
  The a-maximization gives $\e = \frac{-18 + \sqrt{679}}{15}$.
  There is no unitarity violating operators. 
  The central charges are $a \simeq 0.7845$ and $c \simeq 1.0658$.
  The dimensions of the chiral operators are $\Delta(M_{1,-1}) \simeq 1.3885$, $\Delta(u) \simeq 2.7770$.

\paragraph{$Sp(3) \times SU(2)$ theory}
  The central charges of the $Sp(3) \times SU(2)$ theory are given by
    \bea
    a
     =     \frac{25}{12}, ~~~~
    c
     =     \frac{29}{12}, ~~~~
    k_{Sp(3)}
     =     5, ~~~~
    k_{SU(2)}
     =     8
    \eea
  The Coulomb branch operator has dimension $4$.
  
  We first consider the deformation associated to the principal embedding of the $Sp(3)$ part.
  In this case we have $M_{j,-j}$ with charge $(0,2+2j)$ where $j=2,4,6$.
  The 't Hooft anomaly coefficients are
    \bea
    \tr J_+
    &=&    \tr J_+^3 
     =     - 3,   \nonumber \\
    \tr J_-
    &=&   5, ~~~~~
    \tr J_-^3
     =      635,   
               \\
    \tr J_+^2 J_- 
    &=&    35, ~~~~~
    \tr J_+ J_-^2
     =   - 179. \nonumber
    \eea  
  The $a$-maximization gives $\e = \frac{291+ \sqrt{9001}}{480}$, which means that $M_{2,-2}$ violates the unitarity bound. 
  After subtracting and redoing the $a$-maximization, we obtain an $\CN=1$ SCFT with $a \simeq 0.8110$ and $c \simeq 0.9125$.
  The scaling dimensions are $\Delta(u) = \Delta(M_{4,-4}) \simeq  1.1247$ and $\Delta(M_{6,-6}) \simeq 1.6870$.
   
  Then let us consider the principal deformation of $Sp(3) \times SU(2)$.
  In addition to the above $M$, we have $M_{1,-1}'$ with $(J_+, J_-)=(0,4)$.
  The 't Hooft anomaly coefficients are
    \bea
    \tr J_+
    &=&    \tr J_+^3 
     =     - 4,   \nonumber \\
    \tr J_-
    &=&   8, ~~~~~
    \tr J_-^3
     =      614,   
               \\
    \tr J_+^2 J_- 
    &=&    38, ~~~~~
    \tr J_+ J_-^2
     =   - 188. \nonumber
    \eea  
  The $a$-maximization gives $\e = \frac{95+ \sqrt{1195}}{162}$, which means that $M_{2,-2}$ and $M_{2,-2}'$ violate the unitarity bound. 
  By $a$-maximizing again, we obtain an $\CN=1$ SCFT with $a\simeq 0.8002$ and $c \simeq 0.9021$.
  The scaling dimensions are $\Delta(u) = \Delta(M_{4,-4}) \simeq  1.1048$ and $\Delta(M_{6,-6}) \simeq 1.6572$.

\paragraph{$SU(3)$ theory}
  The central charges are given by
    \bea
    a
     =     \frac{71}{24}, ~~~~
    c
     =     \frac{19}{6}, ~~~~
    k_{SU(3)}
     =     14.
    \eea
  The Coulomb branch operator has dimension $6$.
  We have $M_{j,-j}$ with charge $(0,2+2j)$ where $j=1,2$.
  The 't Hooft anomaly coefficients are
    \bea
    \tr J_+
    &=&    \tr J_+^3 
     =     - 2,   \nonumber \\
    \tr J_-
    &=&  -2, ~~~~~
    \tr J_-^3
     =     - 194,   
               \\
    \tr J_+^2 J_- 
    &=&    30, ~~~~~
    \tr J_+ J_-^2
     =   - 34. \nonumber
    \eea  
  The $a$-maximization gives $\e = \frac{2}{3}$, which implies that the dimension of $M_{1,-1}$ is $1$.
  Therefore this operator simply is free and decouple. 
  By subtracting this contribution, we obtain
    \bea
    a
     =     \frac{89}{48}, ~~~
    c
     =     \frac{47}{24}.
    \eea
  This theory has two chiral operators with dimensions $\frac{3}{2}$ and $3$.
  We are not aware of whether this is an $\CN=2$ SCFT or not.

\paragraph{$SU(4)$ theory}
  The central charges of this theory are given by
    \bea
    a
     =     \frac{25}{8}, ~~~~
    c
     =     \frac{7}{2}, ~~~~
    k_{SU(4)}
     =     14.
    \eea
  The Coulomb branch operator has dimension $6$.
  We have $M_{j,-j}$ with charge $(0,2+2j)$ where $j=1,2,3$.
  The 't Hooft anomaly coefficients
    \bea
    \tr J_+
    &=&    \tr J_+^3 
     =     - 3,   \nonumber \\
    \tr J_-
    &=&  -3, ~~~~~
    \tr J_-^3
     =     - 363,   
               \\
    \tr J_+^2 J_- 
    &=&    37, ~~~~~
    \tr J_+ J_-^2
     =   - 83. \nonumber
    \eea  
  The $a$-maximization gives $\e = \frac{3}{4}$, which implies the $M_{1,-1}$ operator decouples along the flow. 
  By subtracting this and doing the $a$-maximization again we obtain the irrational central charges $a \simeq 1.5356$ and $c \simeq 1.6912$.
  These are three chiral operators with dimensions $\Delta(M_{2,-2}) \simeq 1.1086$, $\Delta(M_{3,-3}) \simeq 1.4781$ and $\Delta(u) \simeq 2.2171$.

\paragraph{$Sp(5)$ theory} 
  The central charges are
    \bea
    a
     =     \frac{41}{12}, ~~~~
    c
     =     \frac{49}{12}, ~~~~
    k_{Sp(5)}
     =     7.
    \eea
  The Coulomb branch operator has dimension $6$.
  We have $M_{j,-j}$ with charge $(0,2+2j)$ where $j=2,4,\ldots, 10$.
  The 't Hooft anomaly coefficients
    \bea
    \tr J_+
    &=&    \tr J_+^3 
     =     - 5,   \nonumber \\
    \tr J_-
    &=&   23, ~~~~~
    \tr J_-^3
     =      4973,   
               \\
    \tr J_+^2 J_- 
    &=&    77, ~~~~~
    \tr J_+ J_-^2
     =   - 765. \nonumber
    \eea  
  The $a$-maximization gives $\e = \frac{2121 + \sqrt{147259}}{2814}$.
  The Coulomb branch operator and $M_{j,-j}$ with $j=2,4,6$ are unitarity violating. 
  After subtracting these contributions and $a$-maximizing again, we obtain an $\CN=1$ SCFT with
  $a
          \simeq 0.7399$ and 
   $c
          \simeq 0.8274$.
  The scaling dimensions of the remaining $M_{j,-j}$ are $\Delta(M_{8,-8}) \simeq  1.1990$ and $\Delta(M_{10,-10}) \simeq 1.4988$ respectively.

\section{From conformal SQCD to Argyres-Douglas theory} \label{sec:SQCD}

\subsection{$SU(N)$ with $2N$ flavors to $(A_1, A_{2N-1})$ theory}
  In this section we consider the case where $\CT$ is $\CN=2$ $SU(N)$ gauge theory with $2N$ fundamental hypermultiplets.
  This theory has $SU(2N) \times U(1)$ flavor symmetry.
  The central charges are
    \bea
    a
     =     \frac{7N^2-5}{24},~~~~~
    c
     =     \frac{2N^2 - 1}{6},~~~~~
    k_{SU(N)}
     =     2N.
    \eea    
  Upon coupling the $SU(N)$ adjoint chiral multiplet $M$, and Higgsing via nilpotent vev, 
  the remaining components of $M$ are $M_j$, where $j=1,\ldots, 2N-1$ with charge $(J_+, J_-) = (0, 2 + 2j)$.
  Thus the anomalies are calculated as
    \begin{align}\begin{split}
    &\tr J_+^3
    =    \tr J_+ 
     =     1-2N ,    \\
    &\tr J_- 
    =    2N^2-3 ,   ~~~~
    \tr J_-^3 
    =    16N^4 - 2N^2-3,  \\
    &\tr J_+^2 J_- 
    =    6N^2-3 ,  ~~~~
    \tr J_+ J_-^2
    =    \frac{-32N^3 + 2N + 3}{3}.
    \end{split} \end{align}
  By maximizing the trial central charge, we see various operators violate the unitarity bounds: 
  all the Coulomb branch operators $\tr \phi^i$ ($i=2,3,\ldots,N$)
  and the $M_j$ with $j=1,2,\dots, N-1$.
  As above these will decouple.
  After subtracting the contribution of these fields, we re-do the $a$-maximization. 
  The result is $\epsilon = \frac{3N+1}{3N+3}$. 
  The field $M_N$ has dimension 1, thus decouples, and
  $M_j$ with $j=N+1, \ldots, 2N-1$ has dimensions 
    \bea
    \Delta(M_j)
     = \frac{j+1}{N+1}
    \eea
  This is exactly the operator spectrum of the $(A_1, A_{2N-1})$ theory which we review in section \ref{subsec:AD}.
  Indeed the central charges are calculated, after subtracting the contribution of $M_N$, as
    \bea
    a
     =     \frac{12N^2 - 5 N - 5}{24(N+1)},~~~~~
    c
     =     \frac{3N^2 -N - 1}{6(N+1)}
    \eea
  Note that the unbroken $U(1)_B$ symmetry of $\CT$ is precisely the $U(1)$ of the $(A_1, A_{2N-1})$ theory. 
  Note also that when $N=2$, the $(A_1, A_{2N-1})$ theory has the enhanced $SU(2)$ flavor symmetry.
  Thus we expect in this case that the global symmetry is also enhanced in the IR.
  
\paragraph{Lagrangian for the $(A_1, A_{2N-1})$ theory}
  \begin{table}
  	\centering
  	\begin{tabular}{|c|ccc|}
	         \hline
		 fields & $SU(N)$ & $U(1)_B$ & $(J_+, J_-)$ \\
		 \hline
		 $q$ & $\square$ & $1$ & $(1, -2N+1)$ \\
		 $\tilde{q}$ & $\bar\square$ & $-1$ & $(1, -2N+1)$ \\
		 $\phi$ & adj & $0$ & $(0, 2)$ \\
		 $M_j, (j=1, 2, \ldots, 2N-1)$ & $1$ & $0$ & $(0, 2j+2)$ \\
		 \hline
	\end{tabular}
  \caption{Matter content of the ``Lagrangian description" for the $(A_1, A_{2N-1})$ theory.}
  \label{table:matterA1A2n1}
  \end{table}

  Before the deformation, we have $2N$ quarks/anti-quarks in the fundamental/anti-fundamental representation of the gauge group 
  that has charge $(1, 0)$.
  The adjoint chiral multiplet in the $\CN=2$ vector has charge $(0, 2)$. 
  Then we add the chiral multiplet $M$ transform under the adjoint of $SU(2N)$, which has the charge $(0, 2)$, with the coupling $W=\tr M\mu$.  

  We can easily get an $\CN=1$ theory after giving the nilpotent vev to $M$. 
  The remaining components of the quarks and $M$ fields are given by the ``Fan" associated to the partition $2N \to 2N$ 
  considered in \cite{Agarwal:2014rua}. 
  In the end, we obtain $SU(N)$ gauge theory with one adjoint chiral multiplet $\phi$ with charge $(0,2)$,
  a pair of fundamental and anti-fundamantal chiral multiplets $q$, $\tilde{q}$ with $(1,-2N+1)$, and 
  gauge-singlet chiral multiplets $M_j$ with charge $(0, 2+ 2j)$ with $j=1,2,\ldots,2N-1$. (See the table \ref{table:matterA1A2n1}.)
  The superpotential is given by
    \bea
    W
     =     \sum_{j=1}^{2N-1} M_j (\phi^{2N-1-j} q \tilde{q} ) \ , 
    \eea
    where $\mu_j = \phi^{2N-1-j} q \tilde{q}$ are the remaining components of the moment map $\mu$ of $SU(2N)$ after nilpotent Higgsing.

\subsection{$Sp(N)$ with $2N+2$ flavors to $(A_1, A_{2N})$ theory}
Let us consider the case where $\CT$ is $\CN=2$ $Sp(N)$ gauge theory with $2N+2$ fundamental hypermultiplets ($4N+4$ fundamental half-hypermultiplets).   This theory has the $SO(4N+4)$ flavor symmetry. The central charges are
\be
 a = \frac{1}{24}N(14N+9) , \quad c = \frac{1}{6}N(4N+3) , \quad k_{SO(4N+4)} = 4N . 
\ee
  We now couple a chiral multiplet $M$ transforming in the adjoint representation of $SO(4N+4)$ and give the principal nilpotent vev to $M$. 
  This will break the $SO(4N+4)$ flavor symmetry completely, and the remaining components of $M$ would be $M_j$ with $j=1, 3, \ldots, 4N+1; 2N+1$ having charges $(J_+, J_-)=(0, 2j+2)$. (Note that there are two $M$'s with $j=2N+1$.) Now, the anomalies are given as 
\begin{align}
\begin{split}
&\tr J_+ = \tr J_+^3 =  -2 (N+1), \\
&\tr J_- = 4 N^2+8 N+6, ~~~~
\tr J_-^3 =128 N^4+416 N^3+500 N^2+264 N+54, \\
&\tr J_+^2 J_- = 2 \left(6 N^2+8 N+3\right), ~~~~
\tr J_+ J_-^2 = - \frac{2}{3} \left(64 N^3+144 N^2+107 N+27\right) . 
\end{split}
\end{align}
Given these anomalies we perform $a$-maximization. We find various operators get decoupled along the RG flow. The decoupled operators are all the Coulomb branch operators $\tr\phi^{2i}$ with $i=1, 2, \ldots, N$ and $M_j$ with $j=1, 3, \ldots, 2N+1$ (there are two $M_j$'s with $j=2N+1$) so we are left with $N$ singlet $M_j$ with $j=2N+3, 2N+5, \ldots, 4N+1$. Upon removing the decoupled operators, the anomalies are 
\begin{align}
\begin{split}
&\tr J_+ = \tr J_+^3 = 0, \\
&\tr J_- = -2N, ~~~~
\tr J_-^3 =2N(27+108N+128N^2 + 48N^3), \\
&\tr J_+^2 J_- = 2N(3+4N), ~~~~
\tr J_+ J_-^2 = - 8N(3+7N+4N^2) . 
\end{split}
\end{align}
Maximizing the trial $a$-function, we obtain
\be
 \e = \frac{7+6N}{9+6N} \ , 
\ee
and the central charges
\be
 a = \frac{N(24N+19)}{24(2N+3)} , \quad c = \frac{N(6N+5)}{6(2N+3)} . 
\ee
The central charges are exactly the same as those of the $(A_1, A_{2N})$ theory. 
  The dimensions of the operators $M_{j}$ are given by 
\be
 \Delta(M_j) = \frac{j+1}{2N+3} \quad (j=2N+3, 2N+5, \ldots, 4N+1) \ , 
\ee
which are the same as those of \eqref{eq:A1A2nOps}. When $N=1$, this is the same flow we considered in section \ref{subsec:D4}. 

\paragraph{Lagrangian for the $(A_1, A_{2N})$ theory}
  \begin{table}
  	\centering
  	\begin{tabular}{|c|ccc|}
		 \hline
		 fields & $Sp(N)$ & $(J_+, J_-)$ & $(R_0, \CF)$ \\
		 \hline
		 $q$ & $\square$ & $(1, 0) $ & $(\half, \half) $\\
		 $q'$ & $\square$ & $(1, -4N-2)$ & $(\half(-4N-1), \half(4N+3) )$ \\
		 $\phi$ & adj & $(0, 2)$ & $(1, -1)$ \\
		 $M_j, (j=1, 3, \ldots, 4N+1)$ & $1$ & $(0, 2j+2)$ & $(j+1, -j-1)$ \\
		 $M'_{2N+1}$ & $1$ & $(0, 4N+4)$ & $(2N+2, -2N-2)$ \\
		 \hline
	\end{tabular}
  \caption{Matter content of the ``Lagrangian description" for the $(A_1, A_{2N})$ theory.}
  \label{table:matterA1A2n}
  \end{table}
  
  One can write down the matter content of the deformed $Sp(N)$ SQCD theory that flows to $(A_1, A_{2N})$ theory. 
  In order to see the remaining quarks after nilpotent Higgsing, we note that the fundamental of $SO(4N+4)$ decomposes into 
    \be
   4N+4 \to 
   V_{2N+1} \oplus V_0 , 
    \ee
  under the principal embedding. Therefore we have two fundamental quarks $q, q'$ having charges $(1, 0)$ and $(1, -2N+1)$ from \eqref{chargeshift}. We have the chiral multiplet $\phi$ in the adjoint of $Sp(N)$ and singlet fields $M_j$ with $j=1, 3, \ldots, 4N+4$ and $M'_{2N+1}$. The matter content of the theory is given as in table \ref{table:matterA1A2n}. The superpotential is given by
\be
 W = \phi qq + \sum_{i=1}^{2N+1} M_{2i-1} \left( \phi^{4N+3-2i} q' q' \right) + M'_{2N+1} q q' \ . 
\ee
The terms $\phi^{4N+3-2i} q'q'$ and $qq'$ are the components of the moment map $\mu$ of $SO(4N+4)$ that survive upon Higgsing.

\subsection{The full superconformal index of $(A_1, A_N)$ Argyres-Douglas theory}

Recently, the superconformal index in various limits for the Argyres-Douglas  theory has been computed \cite{Buican:2015ina, Cordova:2015nma, Buican:2015tda,Song:2015wta, Cecotti:2015lab, Cordova:2016uwk}. Here, we compute the full superconformal index of the Argyres-Douglas theory using the gauge theory description we obtained.\footnote{The index computation does not depend on the energy scale of the theory as is well-known, and one can perform the localization at any scale if it is possible. The Lagrangian model we give is not UV complete, but is a theory below the scale specified by a vev which we gave. Thus it is possible to use this Lagrangian to compute the partition function.} 

\paragraph{Generalities}

The superconformal index \cite{Kinney:2005ej, Romelsberger:2005eg} for the $\CN=1$ theory is defined as
\be
 \CI_{\CN=1} (p, q, t; \vec{a}) = \tr (-1)^F p^{j_1 +j_2 + \frac{R}{2}} q^{j_2 - j_1 + \frac{R}{2}} \xi^{F} \prod_i {a_i}^{F_i} \ , 
\ee    
where $(j_1, j_2)$ are the Cartans of the Lorentz group $SU(2)_1 \times SU(2)_2$, and $R$ is the $U(1)_R$ charge, $\CF$ is the global $U(1)_\CF$ charge and $F_i$ are the Cartans for the global symmetries. Here, $R$ can be any candidate $R$-charge, which we pick to be $R_0 = \half (J_+ + J_-)$. Upon finding the superconformal $R$ charge via $a$-maximization, we rescale $\xi \to (pq)^{\frac{\e}{2}} \xi$ to obtain the proper index. 

The superconformal index for a gauge theory can be computed by multiplying contributions from the matter contents and then by integrating over the gauge group. For a chiral multiplet of $R$-charge $r$, and $\CF$-charge $f$, the index is given by
\be
 \CI_{\textrm{chiral}}^{(r, f)} (p, q, \xi; \vec{a})  = \prod_{w_i \in R} \G ((pq)^{\frac{r}{2}} \xi^f \vec{a}^{w_i} ) \ , 
\ee
where $R$ denotes the set of weights in the representation of the flavor symmetry where the chiral multiplet is in. 
The elliptic gamma function is defined as
\be
 \G (z) \equiv \G(z; p, q) = \prod_{m, n=0}^\infty \frac{1 - z^{-1} p^{m+1} q^{n+1} }{1 - z p^m q^n} \ . 
\ee
We also use the standard abbreviated notation $
 \vec{z}^{\vec{w}} \equiv \prod_i z_i^{w_i}$, 
and $f(z^\pm) \equiv f(z^+) f(z^-)$. The vector multiplet contributes to the index by
\be
 \CI_{\textrm{vec}} (p, q) = \k^r \prod_{\vec\a \in \Delta_G} \frac{1}{\G(\vec{z}^{\vec\a})} \ , 
\ee
where $\Delta_{G}$ is the set of all roots of $G$ and $\k = (p; p)(q; q)$. Here $(z; q) = \prod_{m=0}^\infty (1-z q^m)$ is the $q$-Pochhammer symbol. 

The $\CN=2$ index is defined as 
\be
 \CI_{\CN=2} (p, q, t) = \tr (-1)^F p^{j_1 +j_2 + \frac{r}{2}} q^{j_2 - j_1 + \frac{r}{2}} t^{R-\frac{r}{2}} \ , 
\ee
where $R, r$ denote the Cartans for the $SU(2)_R \times U(1)_r$.\footnote{Our $r$ charge is defined in such a way that the Coulomb branch operators have $2\Delta = r$.} The index gets contributions from the states satisfying $\Delta \equiv E - 2j_2 - 2R - r/2 = 0$ where $E$ is the scaling dimension. The fugacities satisfy
\be
 |p| <1 , \quad |q| < 1, \quad |t| < 1, \quad |\frac{pq}{t}| < 1 \ . 
\ee
One can map $\CN=1$ fugacities to $\CN=2$ fugacities by mapping $\xi \to (t(pq)^{-\frac{2}{3}})^\b$ where $\b$ depends on the normalization of the $U(1)_\CF$ charge inside $SU(2)_R \times U(1)_r$. 

Sometimes, it is useful to use the following reparametrization $p = \ft^3 y, q=\ft^3/y, t = \ft^4/v$ to write
\be \label{eq:N2idxtyv}
 \CI_{\CN=2} (\ft, y, v) = \tr (-1)^F \ft^{2(E+j_2)} y^{2j_1} v^{-R+ \frac{r}{2}} \ , 
\ee
with $|\ft| < 1$. This expression makes it easier to expand the index in terms of $\ft$.

\paragraph{$(A_1, A_{2N-1})$ theory}
 Using the $\CN=1$ gauge theory description for the $(A_1, A_{2N-1})$ theory we discussed in this section, it is possible to compute the supersymmetric indices. Here, we compute the full superconformal index and compare against the partial results. 
 
    As we have analyzed, many of the gauge invariant operators of this theory hit the unitarity bound and get decoupled. The decoupled operators are $\tr \phi^{i}$ with $i=2, 3, \ldots, N$ and $M_j$ with $j=1, \ldots, N$. Among the $M_j$ operators, $N-1$ of them with $j=N+1, \ldots, 2N-1$ survives and become the Coulomb branch operators at the end of the RG flow.

Hence, we have the following integral for the index
\begin{align}
 \CI_{\CN=1}^{(A_1, A_{2N-1})} &= \frac{\prod_{j=N+1}^{2N-1} \G \left((pq)^{\frac{j+1}{2}} \xi^{-(j+1)} \right)}{\prod_{i=2}^N \G \left((pq)^{\frac{i}{2}} \xi^{-i}\right)}  \\
  &\quad \times \frac{\k^{N-1}}{N!}  \G\left( (pq)^{\half} \xi^{-1} \right)^{N-1} \oint [d\vec{z}] \prod_{\vec{a} \in \Delta} \frac{\G(\vec{z}^{\vec\a} (pq)^{\half} \xi^{-1})}{\G(\vec{z}^{\vec\a})} 
  \prod_{\vec{w} \in R} \G \left((\vec{z}^{\vec{w}} a)^{\pm} (pq)^{\frac{1-N}{2}} \xi^N \right)  , \nn 
\end{align} 
where $[d\vec{z}] =  \prod_{i=1}^{N-1} \frac{dz_i}{2\pi i z_i}$, $\Delta$ is the set of all roots of $SU(N)$ and $R$ is the set of weights of the fundamental representation of $SU(N)$. The integration contour is given by the unit circle $|z_i|=1$. 
The numerator in the first line comes from the $M_j$ fields that remain coupled in the IR. The denominator comes from the decoupled $\tr \phi^i$ operators. The second line comes from the gauge fields and matter fields $\phi, q, \tilde{q}$. 

The fugacity $\xi$ has to be redefined by $\xi \to \xi' (pq)^{\frac{\e}{2}}$ since $R_{IR} = R_0 +\e \CF$. Furthermore, we map to the $\CN=2$ fugacities by taking $\xi' \to (t (pq)^{-\frac{2}{3}} )^{\frac{1}{N+1}}$. So we take $\xi \to (pq)^{\frac{N-1}{2N+2}} t^{\frac{1}{N+1}}$. Upon this reparametrization, we obtain the following integral
\begin{align} \label{eq:A1A2n1idx}
 \CI_{\CN=2}^{(A_1, A_{2N-1})}  &= \prod_{i=1}^{N-1} \frac{\G\left( (\frac{pq}{t})^{\frac{2N+1-i}{N+1}} \right) }{ \G \left( (\frac{pq}{t})^{\frac{i+1}{N+1}} \right) } \\
 &~ \times \frac{\k^{N-1}}{N!} \G\left( (\frac{pq}{t})^{\frac{1}{N+1}} \right)^{N-1} \oint [d\vec{z}]  \prod_{i \neq j} \frac{\G \left(\frac{z_i}{z_j} (\frac{pq}{t})^{\frac{1}{N+1}} \right)}{\G \left( \frac{z_i}{z_j} \right)}
 \prod_{i=1}^N \G \left(( z_i a)^{\pm} (\frac{pq}{t})^{\frac{1-N}{2N+2}} t^{\half} \right)  . \nn
\end{align}
The integral is over the unit circles, but one has to be careful about the modulus of the integrand. It is most straight-forward to evaluate after reparametrizing the fugacities to $p = \ft^3 y, q = \ft^3/y, t = \ft^4 /v$ and then expand the integrand in $\ft$. We claim this expression gives the full superconformal index of the $(A_1, A_{2N-1})$ theory. In the following, we perform a number of checks against other results in the simplification limits. 

The Coulomb branch limit of the index \cite{Gadde:2011uv} is obtained by taking $p, q, t \to 0$ while $\frac{pq}{t}=u$ fixed. This gives us the integral
\begin{align}
 \CI_{C} (u) = \left( \prod_{i=1}^{N-1} \frac{1}{1-u^{\frac{2N+1-i}{N+1}}} \right) \times \left[ \prod_{i=1}^{N-1} \frac{1-u^{\frac{i+1}{N+1}}}{ 1 - u^{\frac{1}{N+1}}} 
 \frac{1}{N!} \oint [d\vec{z}] \prod_{i \neq j}  \frac{1-z_i/z_j}{1 - u^{\frac{1}{N+1}} z_i/z_j} \right] \ , 
\end{align}
where the expression inside the bracket becomes $1$ upon evaluating the integral. We do not have an analytic proof of this, but we have checked for a number of cases. This indeed agrees with the general expectation that the Coulomb branch is freely generated so that the Coulomb branch index (or the Hilbert series on the Coulomb branch) should be simply given by 
\be
 \CI_{C} (u) = \prod_i \frac{1}{1-u^{\Delta(\CO_i)}} \ , 
\ee
where the product is over the Coulomb branch operators $\CO_i$. See also \cite{Buican:2014qla}. 

When $N=2$, for example, we obtain the following index
\begin{align}
\begin{split}
 \CI_{(A_1, A_3)} &= 1+\ft^{8/3} v^{4/3}- \ft^{11/3} v^{1/3} \chi_2 (y) + \ft^4 v^{-1} \chi_3 (a)   + \ft^{14/3} v^{-2/3}+\ft^{16/3} v^{8/3} \\
 & \quad + \ft^{17/3} v^{4/3} \chi_2 (y) - \ft^6 (\chi_3 (a) +\chi_1 (a) ) - \ft^{19/3} v^{5/3} \chi_2 (y)  \\
 & \quad -\ft^{20/3} v^{1/3} (\chi_3(y) + \chi_1(y)) + \ft^7 v^{-1} \chi_2(y) (\chi_3(a) +\chi_1(a)) +\ft^{22/3} v^{2/3} \\
 & \quad + \ft^{23/3} v^{-2/3} \chi_2 (y) + \ft^8 \left(v^{-2} \chi_5 (a) + v^4 + v \right) + \ft^{25/3} v^{8/3} \chi_2(y) + \ldots  \ , 
\end{split}
\end{align}
and for $N=3$, we obtain
\begin{align}
\begin{split}
 \CI_{(A_1, A_5)} &= 1 + \ft^{5/2}v^{5/4} + \ft^3 v^{3/2} - \ft^{7/2}v^{1/4} \chi_2(y) + \ft^4 (v^{-1} - v^{1/2}\chi_2(y)) \\
  & \quad + \ft^{9/2} v^{-3/4} + \ft^5 (v^{-1/2} + v^{5/2}) + \ft^{11/2} (v^{5/4} \chi_2(y) + v^{11/4}) \\
  & \quad + \ft^6 (-2 + v^{-3/2}(a^3 + a^{-3}) + v^3) + \ldots \ , 
\end{split}
\end{align}
where $\chi_n$ denotes the character for the $n$-dimensional irreducible representation of $SU(2)$. Here the term $\ft^4 v^{-1}$ comes from the conserved current multiplet. We see that for $(A_1, A_3)$ theory, there is a $SU(2)$ current transforming under the adjoint of $SU(2)$ flavor symmetry. For the $(A_1, A_5)$ theory, the coefficient in front of $\ft^4 y^{-1}$ is $1$, which implies there is a $U(1)$ conserved current. 

We find this expression agrees with the index in the Macdonald and Schur limits \cite{Gadde:2011ik, Gadde:2011uv} computed in \cite{Buican:2015ina,Cordova:2015nma,Buican:2015tda, Song:2015wta, Cecotti:2015lab}. Especially, we find that in the Schur limit $t \to q$, the expression becomes independent of $p$. This is consistent with the expectation that our theory preserves $\CN=2$ supersymmetry in the IR.\footnote{We would like to thank Abhijit Gadde for pointing this out to us.}
To be more precise, one can identify from the index that there is indeed a contribution from the $\CN=2$ stress tensor multiplet, which contains the R-symmetry current. 

We would like to point out that it is rather non-trivial to show that the integral \eqref{eq:A1A2n1idx} agrees with other expressions, because they are written in very different manners. Moreover, in the Macdonald limit, our integrand is singular. Therefore it is not so straight-forward to evaluate the simplification limits. For example, since the Higgs branch is given by a simple orbifold $\IC^2/\IZ_N$, the Hall-Littlewood limit of the index ($p \to 0, q \to 0$) is given by (see also \cite{DelZotto:2014kka})
\be
 I_{HL}^{(A_1, A_{2N-1})}(t; a) = \frac{1-t^{N}}{(1-t)(1-t^{\frac{N}{2}} a^N)(1-t^{\frac{N}{2}} a^{-N})} \ .
\ee
It would be interesting to find a proof that this expression agrees with the limit of our integral formula. 

\paragraph{$(A_1, A_{2N})$ theory}
Let us compute the superconformal index for the $(A_1, A_{2N})$ theory using the gauge theory we obtained. Note that the operators $\tr \phi^{2i}$ with $i=1, 2, \ldots, N$ decouples along the RG flow. Among the $M_j's$, the components that remain coupled in the IR are $j=2N+3, 2N+5, \ldots, 4N+1$. From this, we get the index as
\begin{align}
 \CI_{\CN=1}^{(A_1, A_{2N})} &= \left[ \prod_{i=1}^N \frac{ \G \left((pq)^{N+i+1} \xi^{-2(N+i+1)} \right)}{ \G \left((pq)^{i} \xi^{-2i}\right)}\right] \G\left( (pq)^{\half} \xi^{-1} \right)^N  \\
  &\quad \times \frac{\k^N}{2^N N!}  \oint [d\vec{z}] \prod_{\vec{a} \in \Delta} \frac{\G(\vec{z}^{\vec\a} (pq)^{\half} \xi^{-1})}{\G(\vec{z}^{\vec\a})} 
  \prod_{\vec{w} \in R} \G \left( \vec{z}^{\vec{w}} (pq)^{\frac{1}{4}} \xi^{\half} \right) \G \left( \vec{z}^{\vec{w}} (pq)^{\frac{-4N-1}{4}} \xi^{\frac{4N+3}{2}} \right)  , \nn 
\end{align} 
where $[d\vec{z}] =  \prod_{i=1}^{N} \frac{dz_i}{2\pi i z_i}$, $\Delta$ is the set of all roots of $Sp(N)$ and $R$ is the set of all weights in the fundamental representation of $Sp(N)$. When $N=1$, this is the integral formula derived in \cite{Maruyoshi:2016tqk}. 
Now, let us replace the $\CN=1$ fugacities to the $\CN=2$ by substituting $\xi \to (t (pq)^{-\frac{2}{3}})^{\frac{1}{2N+3}}$. Then we get the integral 
\begin{align}
 \CI_{\CN=2}^{(A_1, A_{2N})} &= \left[ \prod_{i=1}^N \frac{ \G \left( (\frac{pq}{t})^{\frac{2(N+i+1)}{2N+3}} \right)}{ \G \left( (\frac{pq}{t})^{\frac{2i}{2N+3}} \right)}\right] \G\left( (\frac{pq}{t})^{\frac{1}{2N+3}} \right)^N  \\
  &\quad \times \frac{\k^N}{2^N N!}  \oint [d\vec{z}] \prod_{\vec{a} \in \Delta} \frac{\G \left(\vec{z}^{\vec\a} (\frac{pq}{t})^{\frac{1}{2N+3}} \right)}{\G(\vec{z}^{\vec\a})} 
  \prod_{\vec{w} \in R } \G \left(\vec{z}^{\vec{w}} (\frac{pq}{t})^{\frac{N+1}{2N+3}} t^{\half} \right) \G \left(\vec{z}^{\vec{w}} (\frac{pq}{t})^{\frac{-N}{2N+3}} t^{\half} \right)  . \nn 
\end{align} 
The integration contour should enclose the poles at $z_i = a^{-1} (\frac{pq}{t})^{\frac{1-N}{2N+2}} t^\half $ but not at $z_i = a^{-1} (\frac{pq}{t})^{-\frac{1-N}{2N+2}} t^{-\half}$. 

The Coulomb branch limit is particularly tractable. In this limit, we obtain
\be
 \CI_C (u) = \left( \prod_{i=1}^N \frac{1}{1- u^{\frac{2(N+i+1)}{2N+3}}} \right)
 \left[ \prod_{i=1}^N \frac{1-u^{\frac{2i}{2N+3}}}{1-u^{\frac{1}{2N+3}}}
 \frac{1}{2^N N!} \oint [d\vec{z}] \prod_{\vec{a} \in \Delta} \frac{1-\vec{z}^{\vec\a}}{1 - u^{\frac{1}{2N+3}} \vec{z}^{\vec\a} } \right] \ , 
\ee
where the terms in the bracket becomes $1$ upon evaluating the integral. This also agrees with the expected result for the Coulomb branch index for the $(A_1, A_{2N})$ theory. 

We have checked that when $N=1, 2$, the leading terms for the Macdonald limit $p \to 0$ and Schur limit $p\to 0, t \to q$ of this integral agrees with the results in \cite{Cecotti:2015lab,Song:2015wta,Cordova:2015nma} computed using different methods. Especially, in the Hall-Littlewood limit, the index becomes trivial since there is no Higgs branch in this theory. 

We note that the Schur index can be written in a simple form by using the Plethystic exponential 
\be
 I_S^{(A_1, A_{2N})}(q) = \PE \left[ \frac{q^2 - q^{2N+2}}{(1-q)(1-q^{2N+3})}\right] \ , 
\ee
which is the same as the vacuum character of the Virasoro minimal model $\CM(2, 2N+3)$. It would be interesting to prove that the Schur limit of our integral formula indeed reproduce this result.  

\paragraph{Checking SUSY enhancement from $\CN=1$ index}
Let us briefly comment on the method to test the enhancement of supersymmetry from computing the $\CN=1$ superconformal index.\footnote{We thank anonymous referee for pointing out such a possibility.} In the section 5.5.1 of the recent paper \cite{Cordova:2016emh}, they list the conserved current multiplets of four-dimensional $\CN=1$ superconformal theory. Once there is a flavor symmetry, we have a conserved current multiplet in the theory. The extra supersymmetry implies that there is extra conserved R-symmetry. But the latter belongs to a different multiplet from the usual conserved current. Therefore, the superconformal index for each multiplets differ. To be concrete, let us compute the $\CN=1$ index 
\be
 I_{\CN=1}(\ft, y) = \tr (-1)^F \ft^{2(E+j_2)} y^{2j_1} \ , 
\ee
for the current multiplets. Note that this trace formulas is the same as the $\CN=2$ version defined in \eqref{eq:N2idxtyv} except for the extra fugacity $v$. The trace is over the states with $E-2j_2 - \frac{3}{2}r=0$. The index for the $\CN=1$ stress tensor is given by 
\be
 I_{\textrm{stress tensor}}(\ft, y) = -\ft^9\left( y+\frac{1}{y} \right), 
 \ee
and an $\CN=1$ flavor current is given by
\be
 I_{\textrm{flavor}}(\ft, y) = -\ft^6 \ . 
\ee
On the other hand, the index for the extended SUSY current is given by
\be \label{eq:idxextSUSY}
 I_{\textrm{extended SUSY}} (\ft, y) = \ft^7 \left(y+\frac{1}{y}\right) + \ft^8. 
\ee
Note that the $\CN=2$ stress tensor gets contributions from all 3 piece we listed above. Therefore, it is possible check whether there is any extended supersymmetry by looking into the $\CN=1$ index and check if there is a contribution of the form given as \eqref{eq:idxextSUSY}. We indeed find such contributions exist for all the cases where we can compute the index.

\section{$T_N$ and $R_{0,N}$ theories of class $\CS$} 
\label{sec:T_N}
  In this section we consider the $\CN=1$ deformation of the $T_N$ and $R_{0,N}$ theories 
  in class $\CS$ \cite{Gaiotto:2009we,Gaiotto:2009hg, Chacaltana:2010ks}.
  We first give a brief review of the class $\CS$ theories. 
  
  The class $\CS$ theory of $A_{N-1}$ type is obtained by the twisted compactification of the six-dimensional $\CN=(2,0)$ theory of $A_{N-1}$ type 
  on a Riemann surface with punctures.
  The four-dimensional theory has $\CN=2$ superconformal symmetry when all the punctures are of regular type
  which is associated to the particular codimension-two half-BPS defect in the six-dimensional theory
  classified by the partition of $N$: $N=\sum_k k n_k$.
  This regular puncture gives the flavor symmetry $S[\prod_k U(n_k)]$ in four dimensions.
  
  For our purpose, let us choose the Riemann surface to be a sphere with three regular punctures. 
  This defines $\CN=2$ SCFTs without exactly marginal deformation. 
  The $T_N$ theory is the one obtained by choosing three maximal punctures given by the partition $n_1=N$, 
  thus has $SU(N)^3$ flavor symmetry.
  The $R_{0,N}$ theory is associated to two maximal punctures and one puncture with the partition $n_1=2$ and $n_{N-2}=1$,
  thus the flavor symmetry is $SU(N)^2 \times SU(2) \times U(1)$.
  It is known that the flavor symmetry in the latter case is enhanced to $SU(2N) \times SU(2)$.
  
  The central charges of the class $\CS$ theories were obtained in \cite{Gaiotto:2009we,Chacaltana:2010ks,Chacaltana:2012zy}.
  Thus it is straightforward to perform the calculation of the deformed $\CN=1$ theory as in the previous sections.
  However let us see more details of the construction which would gain an insight to the deformation procedure.
  
  Starting from the maximal puncture associated to the partition $n_1 = N$, we can get the other type of regular punctures
  by giving the nilpotent vev of the moment map operator $\mu$.
  For example one can get the $R_{0,N}$ theory from the $T_N$ theory by the nilpotent vev of $\mu$ 
  which is the moment map operator of the one of the puncture.
  This higgsing looks similar to our deformation procedure, 
  but a crucial difference is that the latter has an additional adjoint chiral multiplet $M$
  and $M$ is given a nilpotent vev rather than $\mu$.
  This difference can be understood once we consider the $\CN=1$ version of the class $\CS$ theories 
  \cite{Bah:2012dg,Beem:2012yn,Gadde:2013fma, Xie:2013gma}. 
  
  The $\CN=1$ class $\CS$ theory is obtained again by the compactification of the same six-dimensional theory 
  on the Riemann surface but with the different twist which preserve only an $\CN=1$ supersymmetry.
  In the M-theory language, we consider the $N$ M5-branes wrapped on the Riemann surface 
  which is the base space of the two line bundles $\CL_1$ and $\CL_2$.
  The $\CN=1$ condition is that the determinant line bundle $\CL_1 \otimes \CL_2$ has to be equal to the canonical bundle $K_{C_{g, n}}$ of the curve where $C_{g,n}$ is the Riemann surface of genus $g$ and with $n$ punctures.
  By denoting the degrees of $\CL_1$ and $\CL_2$ as $p$ and $q$ respectively, the condition is $p + q = 2g - 2 +n$.
  The puncture could be singular either in the fiber directions of the line bundles. 
  (One could have a puncture where both of them are singular. But we do not consider it here.)
  The theory gets back to $\CN=2$ class $\CS$ if one of the line bundle is trivial and punctures are not singular in this bundle.
  Therefore we notice that there is an additional $\mathbb{Z}_2$ label to the puncture and the pair of pants,
  which we denote as $\sigma_p=\pm1$ and $\sigma_b =\pm1$.
  
  It was found in \cite{Gadde:2013fma} that the four-dimensional description corresponding to the $\sigma_p=-1$ maximal puncture
  attached to the $\sigma_b=+1$ pair of pants is to add the chiral multiplet $M$ transforming in the adjoint representation of the $SU(N)$ flavor symmetry, and the superpotential coupling $\Tr M \mu$.
  The other punctures labelled by general partitions are obtained by giving the corresponding nilpotent vev to $M$.
  At this stage, one could see this is precisely the procedure which we are considering in this paper 
  if we could identify $\CT$ with the class $\CS$ theory
  and the $F$ comes from the regular maximal puncture.
  The principal embedding breaking all the flavor symmetry $F$ corresponds to closing the puncture.

\subsection{Deformation of $T_N$ theory}
\label{subsec:T_N}
  The central charges of the $T_N$ theory is given by
    \bea
    a
     =     \frac{8 N^3-15 N^2-3 N+10}{48}, ~~~
    c
     =     \frac{2 N^3-3 N^2-N+2}{12} , ~~~
    k_{SU(N)}
     =     2N.
    \eea
  It is known \cite{Gaiotto:2009we,Gadde:2011uv,Maruyoshi:2013hja,Beem:2014rza, Lemos:2014lua} that there are Higgs branch operators 
  $\mu^{(A), (B), (C)}$ of dimension 2 that transforms in the adjoint representation of $SU(N)_{A, B, C}$
  and $Q^{(k)}$ transforming in the $(\wedge^k,\wedge^k,\wedge^k)$ representation of $SU(N)_{A, B, C}$ 
  where $\wedge^k$ is the $k$-th anti-symmetric representation of $SU(N)$. 
  The dimension of $Q^{(k)}$ is $k(N-k)$.
  Thus, the $J_+$ and $J_-$ charges of $\mu^{(A), (B), (C)}$ and $Q^{(k)}$ are $(J_+, J_-) = (2,0)$ and $(k(N-k),0)$ respectively.
  There are also the Coulomb branch operators $u_{d,i}$ with dimension $d$, where $d=3,4,\ldots,N$ and $i=1,2,\ldots,d-2$.
  The charges of these operators are $(J_+, J_-)= (0, 2d)$.

\paragraph{Two-punctured sphere}
  Now, let us consider the deformations by adding the chiral multiplet $M$ transforming under the one of the $SU(N)$ flavor symmetry 
  and the superpotential $W = \tr M \mu$.
  In the class $\CS$ language this corresponds to a sphere with $\sigma_b=+1$ and two maximal punctures with $\sigma_p =+1$
  and one maximal puncture with $\sigma_p =-1$.
  Then give nilpotent vev $\rho(\s^+)$ corresponding to the principal embedding $n_N=1$. 
  As mentioned above this corresponds to closing of the puncture with $\sigma_p=-1$, 
  so that it decreases the degree of the normal bundle. 
  This procedure realizes two-punctured sphere with $(p,q) = (1, -1)$. 
  This theory has been already discussed in \cite{Agarwal:2015vla}.

  After the deformation, the shifted charges of the chiral operators $M_{J, -J}$ is $(0, 2J+2)$ where $J=1,2,\ldots,N-1$.
  The 't Hooft anomaly coefficients are calculated by following \eqref{totalanom}:
    \begin{align}
    \begin{split}
    &\tr J_+^3
    =    \tr J_+ 
    =     1-N ,    \\
    &\tr J_-^3
     =     \tr J_-
    =     (1-N)(2N+1),   
               \\
    &\tr J_+^2 J_- 
    =    \frac{(N-1)(4N^2-2N-3)}{3} ,  \\
    &\tr J_+ J_-^2
    =    \frac{(1-N)(4N^2+4N+3)}{3}.
    \end{split}
    \end{align}
  Now, we $a$-maximize to obtain $\e = \frac{1}{3} \sqrt{3+\frac{2}{N}}$. 
  With this value of $\e$, we find that all the $M_{J, -J}$ operators have $R$-charges $R(M_{J, -J}) = (1-\e)(j+1) > \frac{2}{3}$ above the unitarity bound for $N>2$
  and similarly all the Coulomb branch operators have $R$-charge greater than $2/3$. 
  The other Higgs branch operators do not violate unitarity bound because their $J_+$ charge is greater than that of $\mu$. 
  When $N=2$, $T_2$ theory is a free theory in the beginning, so we simply get a theory of free chiral multiplets.

\paragraph{One-punctured sphere}
   Now, let us further close the one of the punctures by considering the same deformation as above. 
   In the class $\CS$ language, the deformed theory corresponds to the one-punctured sphere with normal bundle of bidegree $(1, -2)$. 
   In addition to $M_{J, -J}$, we get another set of singlets $M'_{J, -J}$ by closing another puncture.  
   This gives us the anomaly coefficients:
    \begin{align}
    \begin{split}
    &\tr J_+^3
    =    \tr J_+ 
    =     2(1-N) ,    \\
    &\tr J_-^3
     =     \tr J_-
    =     N(1-N),   
               \\
    &\tr J_+^2 J_- 
    =    \frac{N(N-1)(4N+1)}{3} ,  \\
    &\tr J_+ J_-^2
    =    \frac{2(1-N)(4N^2+4N+3)}{3}.
    \end{split}
    \end{align}
  Upon $a$-maximization, we obtain
    \begin{align}
    \e = \frac{N^2+N+\sqrt{28 N^4+44 N^3+41 N^2+20 N+4}}{9 N^2+6 N+6} \ . 
    \end{align}
  We find that this value of $\e$ makes the singlets of charge $(J_+, J_-)=(0, 4)$ to have $R$-charge below the unitarity bound. Therefore, the operators $M_{1, -1}$ and $M'_{1, -1}$ are decoupled along the RG flow. 

  Now, let us redo $a$-maximization. Removing the decoupled chiral multiplets, we get the anomaly coefficients as
    \begin{align}
    \begin{split}
    &\tr J_+^3
    =    \tr J_+ 
    =     2(1-N) +2,    \\
     &\tr J_-
    =    N(1-N) - 6,    \\
    &\tr J_-^3
    =     N(1-N)- 54,   
               \\
    &\tr J_+^2 J_- 
    =    \frac{N(N-1)(4N+1)}{3} -6 ,  \\
    &\tr J_+ J_-^2
    =    \frac{2(1-N)(4N^2+4N+3)}{3} + 18.
    \end{split}
    \end{align}
  Upon $a$-maximizing again, we obtain
    \be
    \e = \frac{N^3-N-48+\sqrt{28 N^6-12 N^5-19 N^4-210 N^3-27 N^2+300 N+196}}{3 \left(3 N^3-N^2-34\right)} \ . 
    \ee
  The correct $a$-function is given by 
    \begin{align}
    \begin{split}
    a 
    &=    - \frac{3}{64}  \Big(N^3 \left(9 \epsilon ^3-3 \epsilon ^2-9 \epsilon +3\right)+N^2 \left(-3 \epsilon ^3+\epsilon +2\right) \\ 
    & \qquad\qquad +N \left(3 \epsilon ^2+6 \epsilon -1\right)-102 \epsilon ^3+144 \epsilon ^2-62 \epsilon +4\Big) + 2 a_{\textrm{free}} \ , 
    \end{split} \\
    \begin{split}
    c 
    &=     \frac{1}{64} \Big(-9 N^3 \left(3 \epsilon ^3-\epsilon ^2-3 \epsilon +1\right)+N^2 \left(9 \epsilon ^3-5 \epsilon -4\right) \\
    & \qquad\qquad +N \left(-9 \epsilon ^2-12 \epsilon +5\right)+306 \epsilon ^3-432 \epsilon ^2+166 \epsilon -8\Big) + 2 c_{\textrm{free}} \ , 
    \end{split}
    \end{align}
  where $a_{\textrm{free}} = \frac{1}{48}$, $c_{\textrm{free}} = \frac{1}{24}$. 
  For example, when $N=3$, we have
    \be
    a \simeq 0.9512 + 2a_{\textrm{free}} \ , \qquad c \simeq 1.165 + 2c_{\textrm{free}} \ . 
    \ee 

\paragraph{Sphere without puncture}
  Now, let us consider closing all the punctures. The corresponding geometry is given by the normal bundle $\CO(-3) \oplus \CO(1) \to \IP^1$. 
  We get the singlets $M_{J, -J}, M'_{J, -J}, M''_{J, -J}$ with $J=1, \ldots N-1$. 
  The anomaly coefficients are:
    \begin{align}
    \begin{split}
    &\tr J_+^3
    =    \tr J_+ 
    =     3(1-N) ,    \\
    &\tr J_-^3
     =     \tr J_-
    =     N-1,   
               \\
    &\tr J_+^2 J_- 
    =    \frac{(N-1)(4N^2+4N+3)}{3} ,  \\
    &\tr J_+ J_-^2
    =    (1-N)(4N^2+4N+3).
    \end{split}
    \end{align}
  Upon $a$-maximization, we get 
    \be
    \e = \frac{N^2+N+\sqrt{13 N^4+26 N^3+29 N^2+16 N+4}}{6 \left(N^2+N+1\right)} \ . 
    \ee
  This gives the operators $M_{J, -J}, M'_{J, -J}, M''_{J, -J}$ with $J=1$ to have $R$-charges below the unitarity bound, 
  therefore they become free and get decoupled along the RG flow. 

  The rest of the operators do not violate the unitarity bound. We can redo $a$-maximization by subtracting the contribution of the decoupled fields.
  This gives 
    \be
    \e = \frac{N^3-N-36+\sqrt{13 N^6-10 N^4-136 N^3+N^2+176 N+100}}{6 \left(N^3-13\right)}. 
    \ee
  The conformal anomalies are
    \begin{align}
    a &= - \frac{3}{32}  \left(N^3 \left(6 \epsilon ^3-3 \epsilon ^2-6 \epsilon +3\right)+N \left(3 \epsilon ^2+4 \epsilon -1\right)-78 \epsilon ^3+108 \epsilon ^2-46 \epsilon +4\right) + 3 a_{\textrm{free}} ,  \\
   c &= \frac{1}{32} \left(-9 N^3 \left(2 \epsilon ^3-\epsilon ^2-2 \epsilon +1\right)+N \left(-9 \epsilon ^2-8 \epsilon +5\right)+234 \epsilon ^3-324 \epsilon ^2+122 \epsilon -8\right) + 3 c_{\textrm{free}} . 
\end{align}
  For example, when $N=3$, we get
    \be
    a \simeq 0.8731 + 3a_{\textrm{free}} \ , \qquad c \simeq 1.092 + 3c_{\textrm{free}} \ . 
    \ee 

Our result here resolves a puzzle raised in \cite{Bah:2012dg}, where they found the $\CN=1$ SCFT coming from the $N$ M5-branes wrapped on a sphere with normal bundle $\CO(-3) \oplus \CO(1)$ seem to violate the bound on the ratio of central charges $a/c$ when $N=2$. As we have seen in this section, this is due to the fact that there are accidental symmetries (not just for $N=2$, but for general $N \ge 2$) coming from the decoupled operators along the RG flow. Especially when $N=2$, we get a free theory. 

\subsection{Deformation of $R_{0,N}$ theory}
  Let us turn to the $R_{0,N}$ theory.
  This has $SU(2) \times SU(2N)$ flavor symmetry, thus the Higgsing of $SU(2N)$ symmetry does not have class $\CS$ meaning.
  
  The central charges of $R_{0,N}$ is given by
    \bea
    a
     =     \frac{7N^2 - 22}{24}, ~~~
    c
     =     \frac{2N^2-5}{6}, ~~~
    k_{SU(2)}
     =     6, ~~~
    k_{SU(2N)}
     =     2N
    \eea 
  The Higgs branch operator is $\mu$ transforming in $({\rm adj}, \mathbf{1})$ of $SU(2N)\times SU(2)$.
  The charges of this are $(J_+, J_-) = (2,0)$.
  The other Higgs brach operators have $J_+$ charge greater than this.
  Also there are Coulomb branch operators $u_d$ with dimension $d$ where $d=3,4,\ldots, N$.
  Their charges are $(J_+, J_-) = (0,2d)$.

\paragraph{$SU(2N)$ deformation}
  Let us first consider the deformation of $R_{0,N}$ taking $F$ to be $SU(2N)$.
  The remaining singlets have charges $(0, 2J+2)$ with $J=1,2,\ldots, 2N-1$.
  It is easy to calculate the anomalies
    \begin{align}
    \begin{split}
    &\tr J_+^3
    =    \tr J_+ 
    =     - 2N +1,    \\
    &\tr J_-^3
    =     16N^4 -  2N^2 - 5,  \\
    &\tr J_-
    =     2N^2 - 5,   
               \\
    &\tr J_+^2 J_- 
    =    6N^2 - 9 ,  \\
    &\tr J_+ J_-^2
    =    \frac{-32N^3 + 2N + 3}{3}.
    \end{split}
    \end{align}
  A-maximization tells us that all the Coulomb branch operators and $M_{J,-J}$ with $J=1,2,\ldots, N$ get decoupled. 
  By subtracting these contributions and re-maximizing, we get
    \bea
    \e = 
    \frac{9 N^4-15 N^2-6 N+27 + 2 \sqrt{9 N^6+12 N^5+7 N^4+42 N^3-96 N^2-18 N+81}}{3 \left(3 N^4+4 N^3-3 N^2-6
   N+9\right)}.
    \eea
  The central charges are $a \simeq 0.5677+4a_{\textrm{free}}$ and $c \simeq 0.6577+4c_{\textrm{free}}$ for the $N=3$ case.

\paragraph{Full Higgsing}
  We can further break the remaining $SU(2)$ symmetry. We do not repeat the calculation here. 
  We find that the decoupled fields are the same as the above case.
  The central charges of the IR theory are $a \simeq 0.53334 +4a_{\textrm{free}}$ and $c \simeq 0.6681+4c_{\textrm{free}}$ for the $N=3$ case.

\section{Discussion}
\label{sec:conclusion}
  In this paper, we considered the $\CN=1$ deformation of $\CN=2$ SCFTs.
  Among various $\CN=2$ SCFTs, we found the deformation of a particular class of theories flow to the IR fixed point 
  with the enhanced $\CN=2$ supersymmetry. We list the summary of our result in the table \ref{table:results}.
\begin{table}
	\centering
	\begin{tabular}{|c||c|c|c|c|c|}
		\hline
		$\CT$ & $ F $ & $\CN=2$ & Sugawara & $k_F$ bound & $\CT_{IR}[\CT, \rho]$ \\
		\hline \hline
		$(A_1, D_k)$, $(k \ge 4)$ & $SU(2)$ & yes & yes & no & $(A_1, A_{k-1})$ \\
		$(I_{N, Nm+1}, F)$ & $SU(N)$ & yes & yes & no & $(A_{N-1}, A_{Nm+N})$ \\
		$H_1$ & $SU(2)$ & yes & yes & yes & $H_0$ \\
		$H_2$ & $SU(3)$ & yes & yes & yes & $H_0$\\
		$D_4$ & $SO(8)$ & yes & yes & yes & $H_0$\\
		$E_6$ & $E_6$ & yes & yes & yes & $H_0$\\
		$E_7$ & $E_7$ & yes & yes & yes & $H_0$\\
		$E_8$ & $E_8$ & yes & yes & yes & $H_0$\\	
		$SU(N)$ SQCD & $SU(2N)$ & yes & yes & yes & $(A_1, A_{2N-1})$ \\
		$Sp(N)$ SQCD & $SO(4N+4)$ & yes & yes & yes & $(A_1, A_{2N})$ \\
		\hline
		$\CN=4$ $SU(2)$ & $SU(2)$ & no & yes & no & new \\
		$[IV^* , Sp(2) \times U(1)]$ & $Sp(2)$ & no(?) & no & yes & new \\
		$[III^*, SU(2) \times U(1)]$ & $SU(2)$ & no & no & no & new \\
		$[III^*, Sp(3) \times SU(2)]$ & $Sp(3) \times SU(2)$ & no & no & yes & new \\
		$[II^*, SU(3)]$ & $SU(3)$ & no(?) & no & no & new \\
		$[II^*, SU(4)]$ & $SU(4)$ & no & no & no & new \\
		$[II^*, Sp(5)]$ & $Sp(5)$ & no & no & yes & new \\
		$T_N$ & $SU(N)^3$ & no & no & yes & new \\ 
		$R_{0, N}$ & $SU(2N)$ & no & no & yes & new \\
		\hline 
	\end{tabular}
	\caption{Summary of results. Here $F$ denotes the global symmetry that is broken by the principal embedding. (not necessarily the same as the full symmetry of $\CT$) We list whether the deformed theory flows to an $\CN=2$ theory and whether $\CT$ satisfies the Sugawara condition for the central charges of the chiral algebra \cite{Beem:2013sza} and whether the flavor central charge saturates the bound of \cite{Beem:2013sza,Lemos:2015orc}.}
	\label{table:results}
\end{table}

To any $\CN=2$ SCFT $\CT$, there is an associated two-dimensional chiral algebra $\chi[\CT]$ as discussed in \cite{Beem:2013sza}. The central charges for the chiral algebra are given as
 \be
  c_{2d} = -12 c_{4d}, \qquad k_{2d} = -\half k_{4d} \ . 
 \ee
If the two-dimensional Virasoro algebra is given by the Sugawara construction of the affine Lie algebra, the 2d central charge has to be given by $c_{2d} = c_{\textrm{Sugawara}}$, where
 \be
  c_{\textrm{Sugawara}} = \frac{k_{2d} \textrm{dim}F }{k_{2d} + h^\vee} \ , 
 \ee
 where $h^\vee$ is the dual coxeter number of the flavor symmetry group $F$. For a general 4d $\CN=2$ SCFT, $c_{2d} \ge c_{\textrm{Sugawara}}$. 

  From the list of $\CN=2$ SCFTs $\CT$ we considered, it is tempting to conjecture that
   the saturation of the Sugawara bound on the central charges is related to the enhancement of the supersymmetry in the IR.
  An exception to this idea is the $\CN=4$ $SU(2)$ SYM. 
  This is possibly due to the enhanced symmetry of the chiral algebra associated to the $\CN=4$ SYM, 
  where $\chi[\CT]$ has two-dimensional $\CN=4$ supersymmetry. 
  It would be interesting to find a criterion for the $\CT$ to exhibit supersymmetry enhancement at the end of the RG flow. 

 In this paper, we have mostly considered deformations corresponding to the principal embedding. It is possible to consider non-principal embedding as well, which will leave some of the flavor symmetry unbroken. It seems the principal embedding is not the essential condition to ensure $\CN=2$ supersymmetry enhancement, as we have seen in the case of the deformation of the $H_2 = (A_1, D_4)$ theory. A systematic study of the deformation associated to the non-principal embedding is work in progress. 
 
 We have discovered simple $\CN=1$ Lagrangian descriptions for the ``non-Lagrangian" $\CN=2$ Argyres-Douglas theories. We expect this gauge theory description to be a useful tool to understand aspects of the Argyres-Douglas theories. As an application, we computed the full superconformal indices. One observation is that our formula obtained from the gauge theory seems to be very different from the ones obtained in \cite{Buican:2015ina,Buican:2015tda,Song:2015wta}, motivated from the M5-brane realization of the Argyres-Douglas theory \cite{Xie:2012hs}. This may be a hint towards  a dual Lagrangian description that leads us to the same IR fixed point described by the Argyres-Douglas theory.

\acknowledgments 
We would like to thank Prarit Agarwal, Philip Argyres, Ken Intriligator and Yuji Tachikawa for helpful discussions. 
We would also like to thank the hospitality of the Simons Center for Geometry and Physics where this work was initiated during the 2015 Summer Workshop in Mathematics and Physics.
The work of KM is supported by the EPSRC Programme Grant EP/K034456/1 ``New Geometric Structures from String Theory.''
The work of JS is supported in part by the US Department of Energy under UCSD's contract de-sc0009919 and also by Hwa-Ahm foundation.

\appendix
\section{Convention}
\label{sec:app}

\subsection{$\CN=2$ SCFT}
\label{subsec:conventionN=2}
  $\CN=2$ superconformal algebra has $U(1)_r$ and $SU(2)_R$ symmetries.
  We denote by $r$ and $I_3$ the charge of the former and those of the Cartan part of the latter respectively.
  Component fields in a free $\CN=2$ vector multiplet have the following $R$ charges: 
\begin{center}
\begin{tabular}{|c|ccc|} 
\hline
$r~ \backslash ~I_3$ & $\half$ & 0 & $-\half$ \\ 
\hline 
0 & & $A_\mu$ &  \\ 
 1 & $\lambda$ & & $\lambda'$\\
 2 & & $\phi$ & \\
 \hline
\end{tabular}
\end{center} 
and the hypermultiplet has the following charges:
\begin{center}
\begin{tabular}{|c|ccc|}
\hline 
$r ~\backslash ~I_3$ & $\half$ & 0 & $-\half$ \\ 
\hline 
-1 & & $\psi$ &  \\ 
0 & $q$ & & $\tilde{q}^\dagger$\\
1 & & $\tilde{\psi}^\dagger$ & \\
\hline
\end{tabular}
\end{center}

 The 't Hooft anomaly coefficients and the conformal anomalies are related by \cite{Anselmi:1997am}
\be
 \tr R_{\CN=2}^3 = \tr R_{\CN=2} = 48 (a-c) \ , \qquad \tr (R_{N=2} I_a I_b ) = \delta_{ab} (4a-2c) \ . 
\ee
In terms of effective number of hyper/vector multiplet, we can write
\be
 n_h = 4(2a-c) \ ,\qquad n_v = 4(5c-4a) \ . 
\ee
There is a relation between the central charges and the dimensions of the Coulomb branch operators \cite{Shapere:2008zf}:
\be
 2a-c = \frac{1}{4} \sum_{i} (2 \Delta(u_i) - 1) \ . 
\ee

  Another representation of the conformal anomalies are given by \cite{Shapere:2008zf}
\be
 a = \frac{1}{4} R(A) + \frac{1}{6} R(B) + \frac{5r}{24} \ , \qquad c = \frac{1}{3} R(B) + \frac{r}{6} \ , 
\ee
where $r$ is the dimension of the Coulomb branch and 
  \bea
  R(A)
   =     \sum_i \Delta(u_i) - r,
  \eea 
  $R(B)$ is the quantity determined from the Seiberg-Witten curve. 

\paragraph{Central charges of $(A_{N-1}, A_{k-1})$ and $(I_{N,k},F)$ theories}
  Let us focus on the theory $(A_{N-1}, A_{Nm+N})$. 
  The Coulomb branch operators are given in \eqref{dimAA}.
  Thus the $R(A)$ is computed as
    \bea
    R(A)
     =     \frac{(k-1) (N-1) (k (2 N-1)-N-1)}{12 (k+N)}.
    \eea
  where $k=Nm+N+1$.
  Also the $R(B)$ is given by \cite{Xie:2013jc}
    \bea
    R(B)
     =     \frac{(k-1)(N-1)Nk}{4 (N+k)}
    \eea
  These lead to the central charges
    \bea
    a
    &=&    \frac{(m+1)(N-1)N(4(m+1)N^2+4(m+3)N+3)}{48(N (m+2)+1)},
    \nonumber \\
    c
    &=&    \frac{(m+1)(N-1)N(N^2(m+1+N(m+3)+1))}{12(N (m+2)+1)}.
    \eea
  
  Let us then consider the $(I_{N,k},F)$ theory.
  In this case $R(A)$, $R(B)$ and $r$ are given by \cite{Xie:2015rpa,Xie:2016evu}
    \be
   R(A)
    = \sum _{i=2}^N \sum _{j=-i+1}^{\left\lfloor \frac{(i-1) k}{N} - 1\right\rfloor} \left( \frac{i k-j N}{k+N} - 1 \right) \ . 
    \ee
  and 
    \be
    R(B)
     = \frac{1}{4} N(N-1)(N + k - 1)  , \quad 
    r = \frac{1}{2} (N-1)(N+k-1) \ , 
    \ee
  One can further simplify $R(A)$ when $k = Nm+1$ to obtain
\begin{align}
\begin{split}
 R(A) 
  &= \frac{(N-1) (k+N-1) (2N^2 + 2kN - 2N - k -1 )}{12 (k+N)}\ . 
\end{split}
\end{align}
  Thus the central charges $a$ and $c$  are given by
    \bea
    a
    &=&     \frac{\left(N^2-1\right) (k+N-1) (4 k+4 N-1)}{48 (k+N)},
               \nonumber \\
    c
    &=&    \frac{1}{12} (N+k-1)(N^2 - 1).
    \eea
  The flavor central charge is given by
   \bea
   k_{SU(N)} = \frac{2N(N+k-1)}{N+k}. 
   \eea

\subsection{$\CN=1$ SCFT}

Relation between $\CN=1$ $U(1)_R$ and $\CN=2$ $U(1)_r$ charge
\be
 R_{\CN=1} = \frac{1}{3} R_{\CN=2} + \frac{4}{3} I_3 \ . 
\ee

Conformal anomaly for an $\CN=1$ theory is 
\be
 a = \frac{3}{32}  \left( 3 \tr R_{\CN=1}^3 - \tr R_{\CN=1} \right) \ , \qquad c = \frac{1}{32} \left(9\tr R_{\CN=1}^3 - 5 \tr R_{\CN=1} \right)
\ee

In terms of $U(1)_\pm$ in class $\CS$ language, we can write $R$-charges as 
\be
 J_+ = 2 I_3 , \qquad J_- = R_{\CN=2} \ .
\ee
We can also write
\begin{align}
\begin{split}
& \tr J_+ = \tr J_+^3 = 0 \ , \\ 
& \tr J_- = \tr J_-^3 = 48(a-c) \\
& \tr J_+^2 J_- =  8(2a-c)  \ , \\
& \tr J_+ J_-^2 = 0 \ . 
\end{split}
\end{align}
In this language, we write the $R$ charges for $\CN=1$ SCFT as 
\be
R_{IR} = R_0 + \e \CF = \frac{1+\e}{2} J_+ + \frac{1-\e}{2} J_- \ , 
\ee
where we define $R_0 = \half (J_+ + J_-)$ and $\CF = \half (J_+ - J_-)$. The trial $a(\e)$ and $c(\e)$ functions are given by
\be
 a(\e) = \frac{3}{32} \left( 3 \tr R^3  - \tr R \right)
\ee

Flavor central charge
\be
 k_F \delta^{ab} = - 3 \tr R_{\CN=1} T^a T^b = -2 \tr R_{\CN=2} T^a T^b \ , 
\ee
where $k_F$ is normalized so that the hypermultiplet in the fundamental representation has $k_F=1$ and a chiral multiplet has $k_F = \half$. 
The contribution to the 1-loop beta function for the flavor current is given by
\be
 \beta = 3 \Tr R_{\CN=1} T^a T^b . 
\ee

\bibliographystyle{jhep}
\bibliography{ADN1}

\end{document}